\documentclass[12pt]{article}

\usepackage{amsmath}
\usepackage{amssymb}
\usepackage{latexsym}
\usepackage{graphicx}
\usepackage{epsfig}
\usepackage{caption}
\usepackage{subcaption}
\usepackage{float}

\def\beq{\begin{equation}}
\def\eeq{\end{equation}}
\def\bea{\begin{eqnarray}}
\def\eea{\end{eqnarray}}

\begin{document}

\begin{titlepage}

\vspace*{1cm}
\begin{center}
{\bf \Large Hawking Radiation Spectra for Scalar Fields by a \\[1mm] Higher-Dimensional
Schwarzschild-de-Sitter\\[3mm] Black Hole}

\bigskip \bigskip \medskip

{\bf T. Pappas}$^1$, {\bf P. Kanti}$^1$ and {\bf N. Pappas}$^2$

\bigskip
$^1${\it Division of Theoretical Physics, Department of Physics,\\
University of Ioannina, Ioannina GR-45110, Greece}

\bigskip \medskip

$^2${\it Nuclear and Particle Physics Section, Physics Department,\\
National and Kapodistrian University of Athens, Athens GR-15771, Greece}

\bigskip \medskip
{\bf Abstract}
\end{center}
In this work, we study the propagation of scalar fields in the gravitational
background of a higher-dimensional Schwarzschild-de-Sitter black hole as well as on
the projected-on-the-brane 4-dimensional background. The scalar fields have also
a non-minimal coupling to the corresponding, bulk or brane, scalar curvature. 
We perform a comprehensive study by deriving exact numerical results for the 
greybody factors, and study their profile in terms of
particle and spacetime properties. We then proceed to derive the Hawking radiation
spectra for a higher-dimensional Schwarzschild-de-Sitter black hole, and we
study both bulk and brane channels. We demonstrate that the non-minimal field
coupling, that creates an effective
mass term for the fields, suppresses the energy emission rates while the cosmological
constant assumes a dual role. By computing the relative energy rates and
the total emissivity ratio for bulk and brane emission, we demonstrate that the
combined effect of a large number of extra dimensions and value of the field
coupling gives to the bulk channel the clear domination in the bulk-brane energy
balance. 

\end{titlepage}

\setcounter{page}{1}

\setcounter{page}{1}

\section{Introduction}

Black holes in four dimensions have been studied in the context of General Theory of
Relativity, with their spacetime properties and existence criteria being classified in detail.
In the context, however, of the theories with extra spacelike dimensions \cite{ADD,RS}
that were proposed more than fifteen years ago,
all the above have been re-considered. In fact, the first gravitational solution describing a
higher-dimensional, spherically-symmetric black hole formed in the presence of a
cosmological constant was found many decades before the formulation of the
aforementioned theories, and is known as the higher-dimensional Schwarzschild-de-Sitter
or Tangherlini solution \cite{Tangherlini}. 

One of the features that have always attracted the interest of scientists is the emission
of Hawking radiation \cite{Hawking} by black holes: it is the manifestation of a quantum
effect in a curved spacetime, that unfortunately has never been detected so far in the
universe. The idea of the existence of extra spacelike dimensions has given a boost
also to this direction, since extra dimensions may facilitate the creation of mini black
holes at particle colliders and thus the observation of the associated Hawking radiation.
As a result, the derivation of the Hawking radiation spectra from higher-dimensional
black holes has been the topic of a significant number of works - for a partial
list see \cite{KMR}-\cite{Dong} while for a more extensive set of references
one may consult the reviews \cite{Kanti:2004}-\cite{PKEW}.

There have been rather few attempts to determine the form and features of the 
Hawking radiation emission spectra for a higher-dimensional Schwarzschild-de-Sitter
black hole. The first such work appeared in 2005 \cite{KGB}, and provided exact
numerical results for minimally-coupled scalar fields propagating both on the brane
and in the bulk and an analytic study for the lowest partial mode and at the
low-energy regime. Soon afterwards, another analytic study \cite{Harmark} determined
the next-to-leading-order term in the expansion of the greybody factor in terms
of the energy of the scalar particle, again for the lowest partial mode. In 2008,
another numerical study \cite{Wu} addressed the question of the emission of
fields with arbitrary spin from a higher-dimensional Schwarzschild-de-Sitter
black hole. 

A much more recent study \cite{Crispino}, that appeared in 2013, raised the
possibility that the scalar field has also a non-minimal coupling with the scalar
curvature, and provided an analytic formulation of the greybody factor for
all partial modes. However, as this analysis was restricted in the case of a
4-dimensional Schwarzschild-de-Sitter black hole, a work of the present authors
\cite{KPP1} 
appeared a year later containing an analytic study of the higher-dimensional
case that provided analytic expressions for the greybody factors for propagation
of non-minimally coupled scalar fields both in the bulk and on the brane.
The properties of the greybody factors in terms of the particle and spacetime
parameters were studied in detail, however, the validity of the analysis was
limited due to the necessary assumption that both the value of the cosmological
constant and the non-minimal coupling constant had to be small. Recently,
two additional works appeared \cite{Anderson, Sporea} that focused again on the
analytic, low-energy derivation of greybody factors for fields propagating in the
background of a Schwarzschild-de-Sitter black hole.

The aforementioned restriction to the values of the parameters of the model,
involved always in the analytic studies, cannot allow the derivation of the complete
Hawking radiation spectra that need the exact form of the greybody factors
valid for arbitrary values of the parameters. To this end, we return to the study
of non-minimally coupled scalar fields propagating in the background of a
higher-dimensional Schwarzschild-de-Sitter black hole, in order to derive the
exact values of the greybody factors for fields propagating both in the bulk
and on the brane. We will study each case separately, assume in general a 
non-vanishing coupling of the field with the corresponding scalar curvature,
and solve by numerical integration the radial part of the field equation, in terms
of which we may determine the greybody factor. Our previously found analytic
solutions will serve as asymptotic boundary conditions for our numerical analysis.
We will compare the numerical results for the greybody factors with the previously
found analytic ones, and then we will study their profile in terms of the
particle parameters (angular-momentum number and non-minimal coupling
constant) and spacetime properties (number of extra dimensions and value
of cosmological constant).   

Next, we will proceed to derive the differential energy emission rates for Hawking
radiation from a higher-dimensional Schwarzschild-de-Sitter black hole in the
form of scalar fields, both in the bulk and on the brane. Again, the profile of the
emission curves in terms of the parameters of the model will be studied in 
a comprehensive way. Having the complete power spectra at our disposal, we
will also compute the relative energy rates for bulk and brane emission as well
as the total emissivity ratio between the two channels. The amount of the energy
of the black hole that is emitted on our brane, where the 4-dimensional potential
observer lives, has always been an important one; therefore, in the context of this
work also, we will attempt to determine whether the brane domination still persists
or whether it is affected, and maybe overthrown, by factors of this model.

The outline of our paper is as follows: in Section 2, we present the theoretical
framework of our analysis, the gravitational background, and the equations of motion
for the propagation of the scalar fields both in the bulk and on the brane. In
Section 3, we focus on the derivation of the greybody factors for brane and bulk
scalar fields by numerically solving the corresponding radial equations. In Section 4,
we determine the differential energy emission rates, for both types of scalar fields
again, study their profile and finally address the question of the relative emission rates
and the total emissivity ratio. We present our conclusions in Section 5.


\section{The Theoretical Framework}

Let us start by determining first the gravitational background for our analysis. We will
consider the following higher-dimensional gravitational theory 
\beq
S_D=\int d^{4+n}x\, \sqrt{-G}\,\left(\frac{R_D}{2 \kappa^2_D} - \Lambda\right),
\label{action_D}
\eeq
where $D=4+n$ is the dimensionality of spacetime, with $n$ denoting an arbitrary number
of space-like dimensions. Also, $\kappa^2_D=1/M_*^{2+n}$ is the higher-dimensional
gravitational constant, with $M_*$ being the fundamental scale of gravity, and $R_D$
the higher-dimensional Ricci scalar. Finally, $G$ stands for the determinant of the
metric tensor $G_{MN}$ and $\Lambda$ is a positive bulk cosmological constant. 

If we vary the above action with respect to the metric tensor $G_{MN}$, we obtain the
Einstein's field equations that have the form
\beq
R_{MN}-\frac{1}{2}\,G_{MN}\,R_D=-\kappa^2_D G_{MN} \Lambda\,.
\label{field_eqs}
\eeq
The above set of equations admit a spherically-symmetric $(4+n)$-dimensional
solution, known as the Tangherlini solution \cite{Tangherlini}, of the form
\beq
ds^2 = - h(r)\,dt^2 + \frac{dr^2}{h(r)} + r^2 d\Omega_{2+n}^2,
\label{bhmetric}
\eeq
where
\beq
h(r) = 1-\frac{\mu}{r^{n+1}} - \frac{2 \kappa_D^2\,\Lambda r^2}{(n+3) (n+2)}\,,
\label{h-fun}
\eeq
and $d\Omega_{2+n}^2$ is the area of the ($2+n$)-dimensional unit sphere
given by
\begin{equation}
d\Omega_{2+n}^2=d\theta^2_{n+1} + \sin^2\theta_{n+1} \,\biggl(d\theta_n^2 +
\sin^2\theta_n\,\Bigl(\,... + \sin^2\theta_2\,(d\theta_1^2 + \sin^2 \theta_1
\,d\varphi^2)\,...\,\Bigr)\biggr)\,.
\label{unit}
\end{equation}
The above solution describes a higher-dimensional Schwarzschild-de-Sitter spacetime,
with the parameter $\mu$ associated with the black-hole mass $M$ through the relation \cite{MP}
\beq
\mu=\frac{\kappa^2_D M}{(n+2)}\,\frac{\Gamma[(n+3)/2]}{\pi^{(n+3)/2}}\,.
\eeq
Depending on the values of the parameters $M$
and $\Lambda$, the Schwarzschild-de-Sitter spacetime may have two, one or zero
horizons \cite{Molina}, that correspond to the real, positive roots of the equation $h(r)=0$.
Throughout our analysis, we will assume that the values of those parameters are such
so that the spacetime has always two horizons: these will be the black-hole horizon, at
$r=r_h$, and the cosmological horizon, at $r=r_c$; the region of interest will be the area
in between, i.e. $r_h<r<r_c$.

We will now assume that a scalar fileld propagates in the aforementioned 
Schwarz\-schild-de-Sitter spacetime (\ref{bhmetric}). This field may couple to 
gravity either minimally or non-minimally and, accordingly, it will be described by the
action 
\beq
S_\Phi=-\frac{1}{2}\,\int d^{4+n}x \,\sqrt{-G}\left[\xi \Phi^2 R_D
+\partial_M \Phi\,\partial^M \Phi \right]\,.
\label{action-scalar-bulk}
\eeq
In the above, $\xi$ is a positive, constant coupling function: for $\xi=0$, we obtain the case of
minimal coupling, while for $\xi \neq 0$, the scalar field couples to the higher-dimensional
Ricci scalar. The latter quantity may easily be found by contracting the Einstein's field
equations (\ref{field_eqs}) by the inverse metric tensor $G^{MN}$ and is given by the expression
\beq
R_D=\frac{2\,(n+4)}{n+2}\,\kappa^2_D \Lambda\,.
\label{RD}
\eeq
We note that $R_D$ is determined by the bulk cosmological constant as expected, however, it
also carries a dependence on the number of extra spacelike dimensions exhibiting a slight
decrease as $n$ increases.

The variation of $S_\Phi$ with respect to the scalar field leads to its equation of motion that reads
\beq
\frac{1}{\sqrt{-G}}\,\partial_M\left(\sqrt{-G}\,G^{MN}\partial_N \Phi\right)
=\xi R_D\,\Phi\,. 
\label{field-eq-bulk}
\eeq
As is usual, we will assume that the effect of the propagation of this massless scalar field 
on the spacetime background is negligible, and thus Eq. (\ref{field-eq-bulk}) will
be solved for a fixed background given by (\ref{bhmetric}). To this end, we assume a
factorized ansatz for the scalar field of the form
\beq \Phi(t,r,\theta_i,\varphi) = e^{-i\omega t}R(r)\,\tilde Y(\theta_i,\varphi)\,,
\eeq
where $\tilde Y(\theta_i,\varphi)$ are the hyperspherical harmonics \cite{Muller}. By
using the eigenvalue equation of the latter, we may easily decouple the radial part of
Eq. (\ref{field-eq-bulk}) from its angular part ending up with the radial equation \cite{KPP1}
\beq
\frac{1}{r^{n+2}}\,\frac{d \,}{dr} \biggl(hr^{n+2}\,\frac{d R}{dr}\,\biggr) +
\biggl[\frac{\omega^2}{h} -\frac{l(l+n+1)}{r^2}-\xi R_D\biggr] R=0\,.
\label{radial_bulk}
\eeq
In \cite{KPP1}, Eq. (\ref{radial_bulk}) was also written in the form of a Schr\"odinger-like
equation with the effective potential `felt' by the bulk scalar field having the explicit form
\beq
V_{\rm eff}^{\rm bulk}=h(r) \left\{\frac{(2l+n+1)^2-1}{4r^2}+
\kappa^2_D \Lambda\,(n+4) \left[\frac{2\xi}{(n+2)}-\frac{1}{2(n+3)}\right]
 + \frac{(n+2)^2 \mu}{4r^{n+3}}\right\}\,.
 \label{pot_bulk}
\eeq
The potential bears a dependence on both spacetime and particle properties.
Its detailed study in \cite{KPP1} revealed that it has the form of a barrier
whose height increases with the number of extra dimensions $n$ of spacetime
as well as with the angular-momentum number $l$ and coupling parameter
$\xi$ of the scalar field. The profile of the potential in terms of the bulk
cosmological constant is more subtle: the height of the barrier decreases
with $\Lambda$ for vanishing or small values of $\xi$, but it increases for
large values of $\xi$. Independently of the values of the aforementioned
parameters, the effective potential always vanishes at the location of the two horizons,
$r_h$ and $r_c$, a feature that allows us to compute the probability for the
emission of these scalar fields by the black hole, i.e. the so-called greybody factors,
by using the asymptotic, free-wave solutions in these two regimes.

The case of a scalar field that is restricted to live on the brane must be studied
separately. Such a field propagates strictly in a 4-dimensional gravitational
background, that is the projection of the line-element (\ref{bhmetric}) on the
brane and follows by fixing the values of the extra angular coordinates,
$\theta_i=\pi/2$, for $i=2, ..., n+1$. Then, we obtain 
\beq
ds^2_4 = - h(r)\,dt^2 + \frac{dr^2}{h(r)} + r^2\,(d\theta^2 + \sin^2\theta\,
d\varphi^2)\,,
\label{metric_brane}
\eeq
where the metric function $h(r)$ is still given by Eq. (\ref{h-fun}). A brane scalar
field that is coupled to gravity either minimally or non-minimally will now be
described by the action functional
\beq
S_4=\frac{1}{2}\int d^4x \,\sqrt{-g}\left[\left(1-\xi \Phi^2\right) R_4
- \,\partial_\mu \Phi\,\partial^\mu \Phi \right].
\label{action4}
\eeq
In the above expression, $g_{\mu\nu}$ is the projected metric tensor on the
brane defined in Eq. (\ref{metric_brane}), and $R_4$ the corresponding brane
curvature found to be 
\beq
R_4=\frac{24 \kappa_D^2 \Lambda}{(n+2) (n+3)} + \frac{n(n-1)\mu}{r^{n+3}}\,.
\label{R4}
\eeq
The brane scalar curvature also carries a dependence on the spacetime parameters
but its expression is distinctly different from the bulk one (\ref{RD}).

The variation of $S_4$ with respect to the brane scalar field $\Phi$ now leads to the
equation of motion
\beq
\frac{1}{\sqrt{-g}}\,\partial_\mu\left(\sqrt{-g}\,g^{\mu\nu}\partial_\nu \Phi\right)
=\xi R_4\,\Phi\,. 
\label{field-eq-brane}
\eeq
We will assume again a factorized ansatz for the field, namely
\begin{equation}
\Phi(t,r,\theta,\varphi)= e^{-i\omega t}\,R(r)\,Y(\theta,\varphi)\,,
\label{facto}
\end{equation}
where $Y(\theta,\varphi)$ are now the usual scalar spherical harmonics. Using the
above factorization and the eigenvalue equation of $Y(\theta,\varphi)$, we obtain the
following decoupled radial equation for the function $R(r)$ \cite{KPP1}
\beq
\frac{1}{r^2}\,\frac{d \,}{dr} \biggl(hr^2\,\frac{d R}{dr}\,\biggr) +
\biggl[\frac{\omega^2}{h} -\frac{l(l+1)}{r^2}-\xi R_4\biggr] R=0\,.
\label{radial_brane}
\eeq
By turning the above equation into a Schr\"odinger-like form, the brane
effective potential is found to be 
\beq
V_{\rm eff}^{\rm brane}=h(r) \left\{\frac{l(l+1)}{r^2}+
\frac{4\kappa^2_D \Lambda(6\xi-1)}{(n+2)(n+3)}\,
 + \frac{\mu}{r^{n+3}}\left[\,(n+1)+\xi n(n-1)\,\right]\right\}\,.
 \label{pot_brane}
\eeq
The above expression shows a similar behaviour, in terms of the spacetime 
and scalar field properties, as in the case of the bulk potential but of different
magnitude at times. lt vanishes again at the two horizons, $r_h$ and $r_c$, due
to the vanishing of the metric function $h(r)$; this will allow us again to
study the emission of brane scalar fields by the black-hole background
(\ref{metric_brane}). Note that the mass parameter $\mu$ can be eliminated
from both our bulk and brane analysis by using the black-hole horizon equation
$h(r_h)=0$; then, we obtain
\beq
\mu=r_h^{n+1}\left(1-\frac{2\kappa^2_D \Lambda r_h^2}{(n+2)(n+3)}\right).
\label{mu}
\eeq
If we then fix the value of the black-hole horizon, i.e. at $r_h=1$, we may
investigate the emission problem of bulk and brane scalar fields in terms
of the number of extra dimensions $n$, cosmological constant $\Lambda$,
angular-momentum number $l$ and coupling constant $\xi$ of the scalar
field.


\section{Greybody Factors} 

In this section, we will determine the greybody factors, i.e. the probability that scalar
particles produced near the black-hole horizon will overcome the effective potential barrier
and propagate away from the black hole. The analysis must be performed separately for
brane and bulk scalar fields since, as we saw, they `see' a different gravitational background
and obey different equations of motion. We will start from the study of scalar fields on
the brane since it is this channel of radiation that may be observed, given that the
potential observers are also restricted to live on the brane. Subsequently, we will turn to
the bulk and perform a similar analysis for higher-dimensional scalar fields.


\subsection{Transmission of Scalar Particles on the Brane} 

The transmission probability for scalar particles propagating in the projected-on-the-brane
black-hole background (\ref{metric_brane}) may be found in terms of the radial function
$R(r)$. To this end, we need to solve the radial scalar equation (\ref{radial_brane}).
Unfortunately, this equation cannot be solved analytically over the whole radial regime
even in the absence of the non-minimal coupling parameter $\xi$ or the cosmological
constant $\Lambda$. 

In a previous work of ours \cite{KPP1}, Eq. (\ref{radial_brane}) was solved analytically
by using an approximate method based on the smooth matching of the asymptotic
solutions found by solving the radial equation close to the black-hole and cosmological
horizons. Our analysis was a comprehensive one, being valid for all partial modes,
labeled by the angular-momentum number $l$, and taking into account the effect of
$\Lambda$ at both asymptotic regimes. We will now briefly present the main results
of our analytic approach as these will be used either as boundary conditions or
as checking points for our exact numerical analysis. 

The radial equation (\ref{radial_brane}) was first solved in the radial regime close
to the black-hole horizon. By using the coordinate transformation
\beq
r \rightarrow f(r) = \frac{h(r)}{1- \tilde \Lambda r^2}\,,
\label{newco-f}
\eeq
where $\tilde \Lambda \equiv 2 \kappa_D^2 \Lambda/(n+2)(n+3)$, the aforementioned equation
takes the form of a hypergeometric differential equation. Applying the boundary condition that
only incoming modes are allowed in the region just outside the black-hole horizon, we obtain the
general solution 
\begin{eqnarray}
&& \hspace*{-1cm}R_{BH}(f)=A_1 f^{\alpha_1}\,(1-f)^{\beta_1}\,F(a_1,b_1,c_1;f),
\label{BH-gen}
\end{eqnarray}
where $A_{1}$ is an arbitrary integration constant, while the hypergeometric indices $(a_1,b_1,c_1)$ 
and the power coefficient $\beta_1$ are defined in terns of both particle and spacetime parameters
-- for more information on this, the interested reader may consult our previous work \cite{KPP1}.
Here, we give only the expression of the remaining power coefficient, namely
\beq
\alpha_1 = -\frac{i \omega r_h}{A_h}\,, 
\label{sol-albe-rh}
\end{equation}
as this will be of use shortly. 
In the above, $A(r)=(n+1)-(n+3)\,\tilde\Lambda r^2$ and $A_h=A(r=r_h)$. 

Near the cosmological horizon, that for a small cosmological constant is located far away
from the black-hole horizon, we used instead the simplified metric function
$h(r) \simeq 1- \tilde \Lambda r^2$ as the new radial variable. Then, the radial equation
(\ref{radial_brane}) in the area close to $r=r_c$ took again the form of a hypergeometric
differential equation with solution
\begin{eqnarray}
&& \hspace*{-1cm}R_{C}(h)=B_1 \,h^{\alpha_2}\,(1-h)^{\beta_2}\,X(a_2,b_2,c_2;h)
\nonumber \\[1mm] && \hspace*{2cm} +\,
B_2\,h^{-\alpha_2}\,(1-h)^{\beta_2}\,X(a_2-c_2+1,b_2-c_2+1,2-c_2;h)\,,
\label{FF2-dS}
\end{eqnarray}
where $B_{1,2}$ are also arbitrary constants, and the hypergeometric indices and power
coefficients are given again in terms of the parameters of the model. 

The matching of the two asymptotic solutions (\ref{BH-gen}) and (\ref{FF2-dS})
at an intermediate radial regime ensures the existence of a complete solution
in the area between the two horizons. As shown in \cite{KPP1}, a smooth matching
takes place under the assumption that both the cosmological constant and the
non-minimal coupling constant remain small. The amplitudes of the incoming and
outgoing wave at the cosmological horizon define the greybody factor for the brane
scalar fields; these are found to be given by the integration constants $B_1$ and $B_2$,
respectively, and thus the greybody factor, or transmission probability, may be written as 
\beq
|A|^2=1-\left|\frac{B_2}{B_1}\right|^2\,.
\label{greybody}
\end{equation}
The ratio $B_2/B_1$ follows from the matching of the two asymptotic solutions and is
expressed as an analytical, but complicated, expression of all the aforementioned indices,
coefficients and parameters of the model \cite{KPP1}. For the case of minimal coupling
($\xi=0$), and for the lower partial mode ($l=0$), we have confirmed that, in the limit
$\omega \rightarrow 0$, the greybody factor takes the simple form
\beq
|A^2|=\frac{4 r_h^2r_c^2}{(r_c^2+r_h^2)^2}+ {\cal O}(\omega)\,,
\label{geom_brane}
\eeq
in terms of the black-hole and cosmological horizons, and in agreement with the results
produced in \cite{KGB,Harmark,Crispino}. If we assume instead that $\xi \neq 0$, then the
low-energy limit of the greybody factors for all partial modes is of order ${\cal O}(\omega r_h)^2$
\cite{KPP1, Crispino} and therefore vanishes for $\omega \rightarrow 0$. We may thus
conclude that the non-vanishing, geometric limit of $|A^2|$ in the low-energy regime is
a characteristic of only free, massless scalar particles in a Schwarzschild-de-Sitter spacetime,
both four- and higher-dimensional; as soon as we turn on the non-minimal
coupling, this feature disappears due to the additional interaction with gravity or,
equivalently, due to the existence of an effective mass-term for the scalar field as follows
from Eq. (\ref{field-eq-brane}). 

The approximate method described above allowed us to find an expression for the greybody
factor fully and explicitly determined by the parameters of system. It also proved to be very
satisfying and trustworthy: not only did it reproduce the correct low-energy geometric limit
but produced, for most choices of parameters, smooth curves over the entire energy regime,
in contrast with other analytic results that usually break down already from the intermediate
energy regime. That was due to the fact that we avoided making any unnecessary simplifications
or approximations regarding the parameters - not even the energy of the emitted particles -
or the mathematical expressions involved. Our only assumptions were the smallness of the
cosmological constant $\Lambda$ and the coupling constant $\xi$. However, as is usual the
case when using this approximate matching technique, deviations from the exact behaviour
are indeed expected to appear when we move beyond the low-energy regime. As a result,
in order to be able to compute the complete energy emission spectra for arbitrary choices
of our parameters, we should use an exact numerical technique for the integration of
Eq. (\ref{radial_brane}) and the derivation of the greybody factor. 

We should first determine the boundary conditions for our numerical analysis. Near the
black-hole horizon, where $r \rightarrow r_h$ and $f \rightarrow 0$, the corresponding
solution (\ref{BH-gen}), takes the form
\beq
R_{BH} \simeq A_1\,f^{\alpha_1} = A_1\,e^{-i(\omega r_h/A_h)\,\ln f}\,.
\label{BH-exp}
\eeq
The above describes indeed an ingoing wave at the black-hole horizon, and, given that the
arbitrary constant $A_1$ carries no physical significance, we may normalize it to unity
by setting 
\beq
R_{BH}(r_h)=1\,. \label{R_BH_num}
\eeq
For the first derivative, we obtain
\beq 
\frac{dR_{BH}}{dr}\biggr|_{r_h}=A_1 f^{-i (\omega r_h/A_h)\,\ln f}\,\biggl(-\frac{i \omega r_h}{A_h}\biggl)
\frac{A(r) (1-f)}{h(r)\,r}\biggl|_{r=r_h} \simeq -\frac{i \omega}{h(r)}\,,
\label{dR_BH_num}
\eeq
where, as before, $A(r)=(n+1)-(n+3)\,\tilde \Lambda r^2$ and where we have used
Eq. (\ref{R_BH_num}) after we performed the derivative.

Although in our analytic approach, the simplified variable $h(r) \simeq 1-\tilde \Lambda r^2$
was used in the regime close to the cosmological horizon, in our numerical analysis, we will keep
the $f$-coordinate (\ref{newco-f}) as this takes into account the full effect of the black-hole
mass and the cosmological constant. Then, following a similar analysis to the one close to the
black-hole horizon, the asymptotic solution as $r \rightarrow r_c$ and $f \rightarrow 0$ is
written as
\beq 
R_C \simeq B_1\,f^{\alpha_2} + B_2\,f^{-\alpha_2} = B_1\,e^{-i (\omega r_c/A_c)\ln f} + 
B_2\,e^{i (\omega r_c/A_c)\ln f} \,, \label{CO-exp}
\eeq
where now $\alpha_2=-i \omega r_c/A_c$.
Since $f$ decreases as the cosmological horizon is approached and $A_c<0$ (as we will
see in Section 4), the term proportional to $B_1$ describes again the ingoing wave and the
term proportional to $B_2$ the outgoing wave. Therefore, the greybody factor is given again
by the expression (\ref{greybody}).

Our numerical integration starts from the black-hole horizon, i.e. from $r=r_h+\epsilon$,
where $\epsilon$ is a small positive number in the range $10^{-6}-10^{-4}$, using the
boundary conditions
(\ref{R_BH_num})-(\ref{dR_BH_num}). Note that both the value of $\epsilon$ and the integration
step are appropriately chosen so that the numerical results are stable. Our numerical
integration proceeds towards the cosmological horizon; there, the values of
the multiplying coefficients $B_1$ and $B_2$ are extracted by using the relations
\beq
B_1=\frac{1}{2}\,e^{i (\omega r_c/A_c)\,\ln f}\,\biggl[R_C(r)+\frac{i A_c h r}{\omega r_c A(r) (1-f)}\,
\frac{d R_C}{dr}\biggr]\,, \label{incoming}
\eeq
\beq
B_2=\frac{1}{2}\,e^{-i (\omega r_c/A_c)\,\ln f}\,\biggl[R_C(r)-\frac{i A_c h r}{\omega r_c A(r) (1-f)}\,
\frac{dR_C}{dr}\biggr]\,. \label{outgoing}
\eeq

\begin{figure}[t!]
    \centering
    \begin{subfigure}[ht]{0.47\textwidth}
        \includegraphics[width=\textwidth]{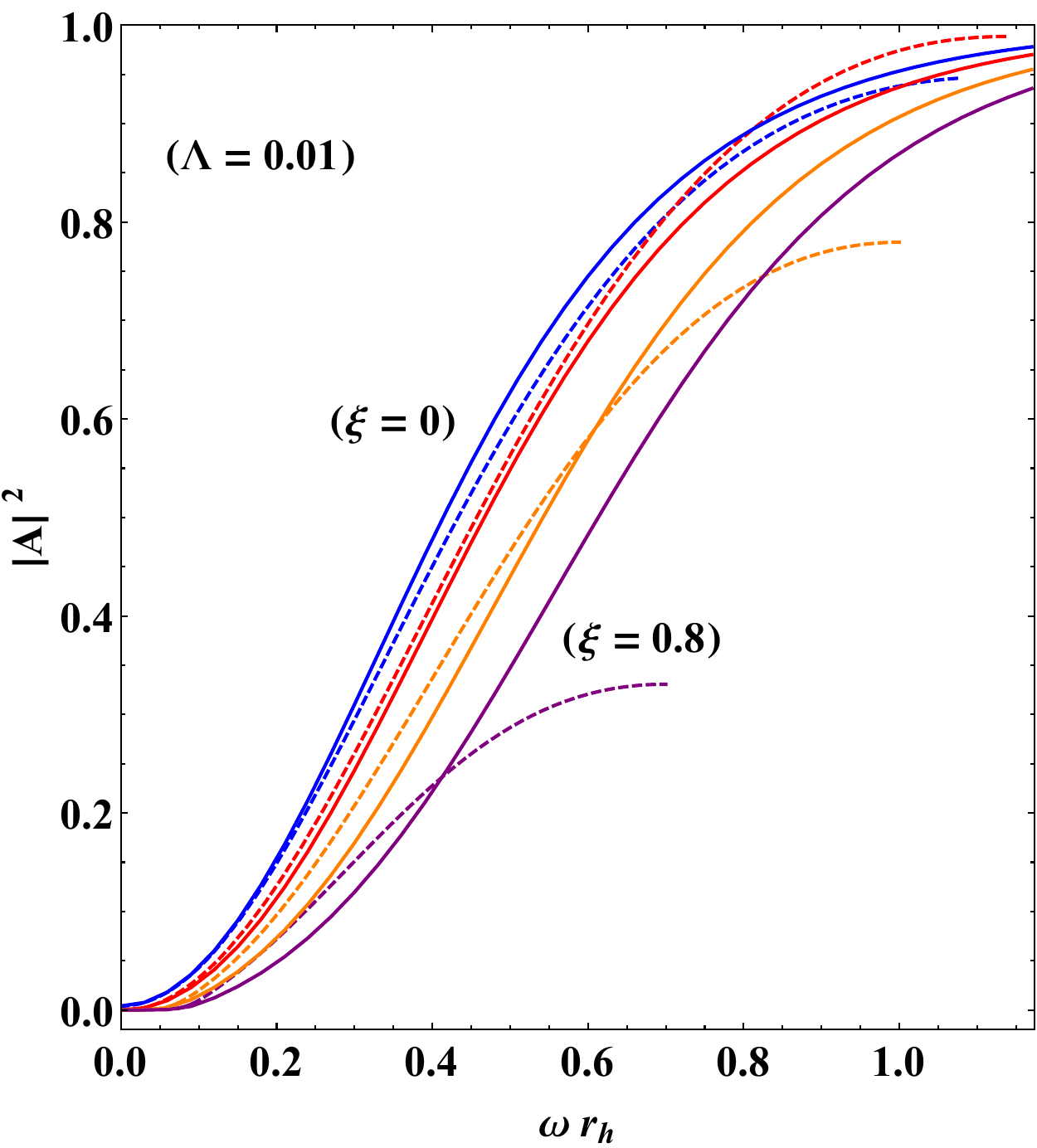}
        \caption{}
        \label{anal_vs_num_br_var_xi}
    \end{subfigure}
    ~ 
    \begin{subfigure}[ht]{0.47\textwidth}
        \includegraphics[width=\textwidth]{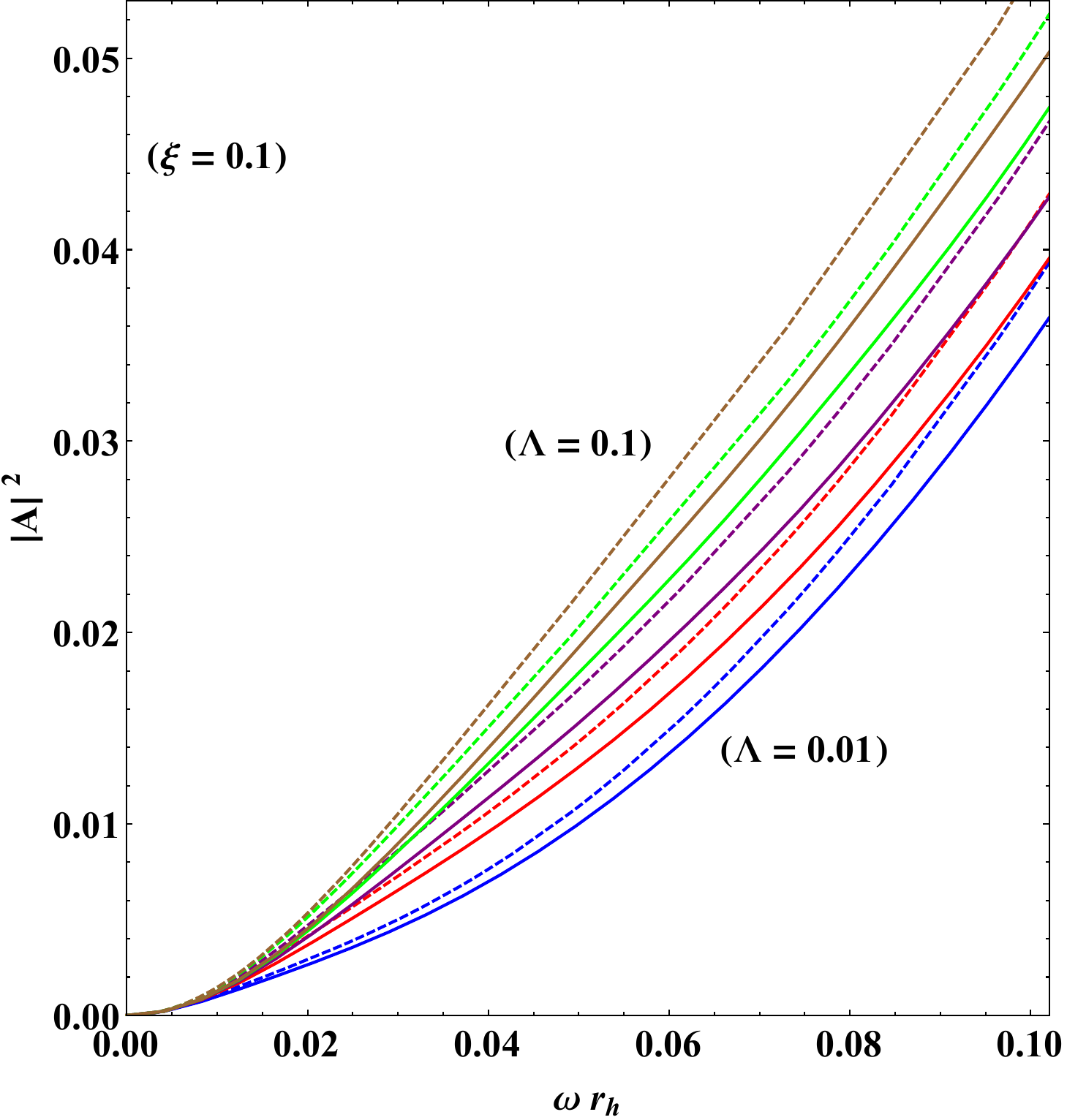}
        \caption{}
        \label{anal_vs_num_br_var_L}
    \end{subfigure}
    \caption{(colour online). Greybody factors for brane scalar fields. Analytical (dashed curves)
and numerical results (solid curves) for $l = 0 , n = 2$ with (a) $\Lambda = 0.01$ (in units of
$r_h^{-2}$) for variable
(top to bottom) $\xi = 0, 0.2, 0.5, 0.8$ and (b) $\xi = 0.1$ for variable (bottom to top)
$\Lambda = 0.01, 0.03, 0.05, 0.08, 0.1$.  }\label{anal_vs_num_br_xi_L}
\end{figure}

Having developed our numerical integration technique, we now have the opportunity to check the
accuracy of the analytic expressions for the greybody factors derived in our previous work \cite{KPP1}. 
To this end, in Fig. \ref{anal_vs_num_br_xi_L}(a,b) we plot the numerical (solid curves) and analytic
results  (dashed curves) for the greybody factors for brane scalar fields
for variable coupling constant $\xi$ and cosmological constant $\Lambda$, respectively.
As expected, the two sets of results are in excellent agreement in the low-energy regime while
for larger values of $\omega r_h$ the analytic results deviate from the exact, numerical ones.
The deviation depends strongly on the values of $\xi$ and $\Lambda$: when both 
parameters are kept small, the deviation is limited and remains so throughout the energy regime;
however, as any of these two parameters increases, the deviation becomes significant and
the range of agreement becomes gradually smaller. The latter was anticipated by the fact that our
analytic approach was valid only for small values of the parameters $\xi$ and $\Lambda$.

We may now use our numerical technique to derive the exact form of the greybody factor for brane
scalar fields, that will be valid over the entire energy regime and for arbitrary values of the
particle and spacetime parameters. Having the exact greybody factors at our disposal is also a
prerequisite for determining the power spectra for the emission of brane scalar fields by a
higher-dimensional Schwarzschild-de-Sitter black hole, a task that we undertake in the following
section.

\begin{figure}[t!]
    \centering
    \begin{subfigure}[ht]{0.47\textwidth}
        \includegraphics[width=\textwidth]{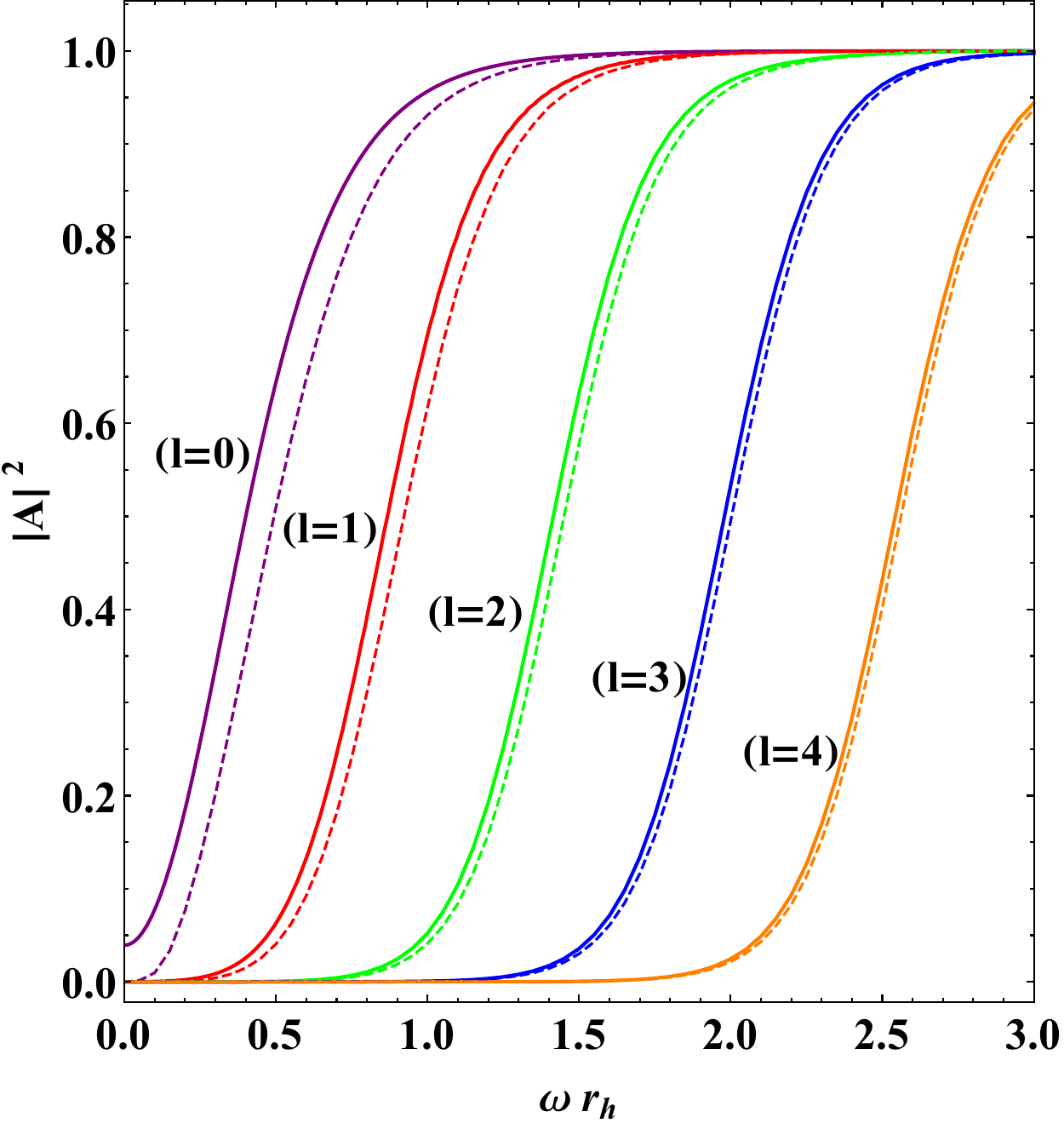}
        \caption{}
        \label{brane_l}
    \end{subfigure}
    ~ 
    \begin{subfigure}[ht]{0.46\textwidth}
        \includegraphics[width=\textwidth]{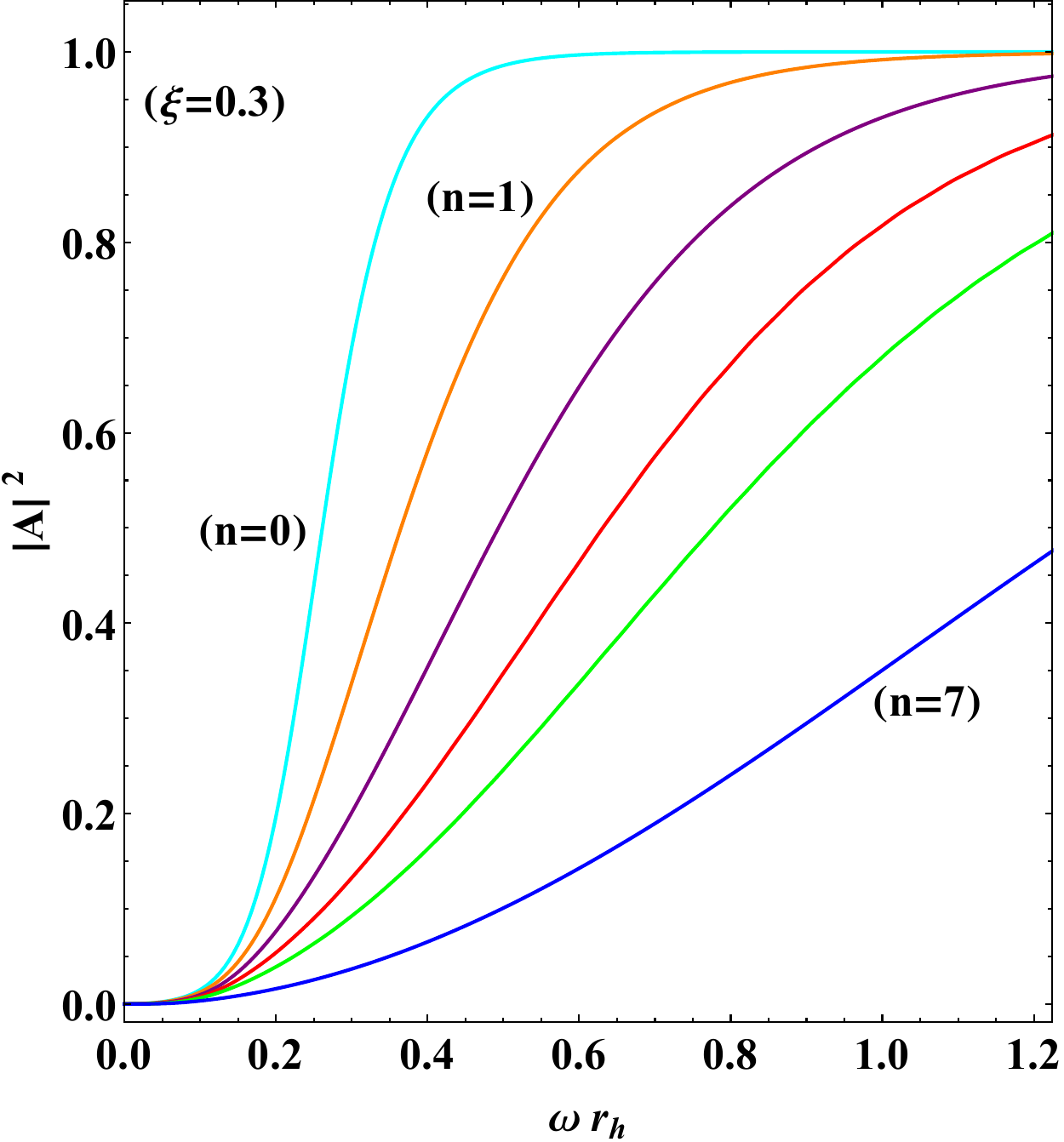}
        \caption{}
        \label{brane_n}
    \end{subfigure}
    \caption{(color online). Greybody factors for brane scalar fields for $\Lambda = 0.1$ and 
(a) $n = 2$ for variable $l = 0,1,2,3,4$ and $\xi = 0$ (solid curves) or $\xi = 0.3$ (dashed
curves); (b) $l = 0 , \xi = 0.3$ for variable (top to bottom) $n = 0, 1, 2, 3, 4, 7$. }
\label{br_l_and_n}
\end{figure}

In Fig. \ref{br_l_and_n}(a), we plot the behavior of the greybody factor under the variation of
the angular momentum number $l$ of the field. Clearly, it is the lowest partial mode ($l=0$)
that has the most enhanced greybody factor while higher-field modes have their greybody
factors suppressed as $l$ increases. This behavior is expected since the background's spherical
symmetry favors the emission of modes with the same type of symmetry. The plot depicts also
the dependence of the greybody factor for minimal and non-minimal coupling: in the case of
the $l = 0$ mode, the non-vanishing asymptotic low-energy limit (\ref{geom_brane}) is
recovered for $\xi = 0$, while for non-vanishing $\xi$, this asymptotic value vanishes. In all
cases, the effect of the coupling constant is to suppress the greybody factor throughout the
energy regime. This effect is more prominent for the lower partial modes and is almost
entirely eliminated for $l \geq 4$. The above behaviour is in excellent agreement with the
analytic one found in \cite{KPP1} for small values of $l$ and in the low-energy regime;
however, as either parameter increases the analytic results suffer from the appearance of poles
that occasionally lead to the abrupt termination of the greybody factor curve. The exact
numerical analysis overcomes this obstacle and provides us with complete, smooth curves.

In Fig. \ref{br_l_and_n}(b), we depict the dependence of the greybody factor on the number
of extra spacelike dimensions $n$. In various previous works, it has been noted that $n$
causes a suppression of the greybody factor for brane scalar fields over the entire energy
regime, and we recover the same behaviour here.  Comparison of the exact numerical results
with the analytic ones of \cite{KPP1} reveal again a very good agreement at the low-energy
regime. As either the energy parameter or the number of dimensions increases, the range
of agreement decreases, and the analytic curves, due again to the existence of poles, tend 
to lie ``lower'' than the exact numerical ones, a behaviour that is found to be common in
all subsequent plots.  

\begin{figure}[t]
  \begin{center}
\mbox{\includegraphics[width = 0.52 \textwidth] {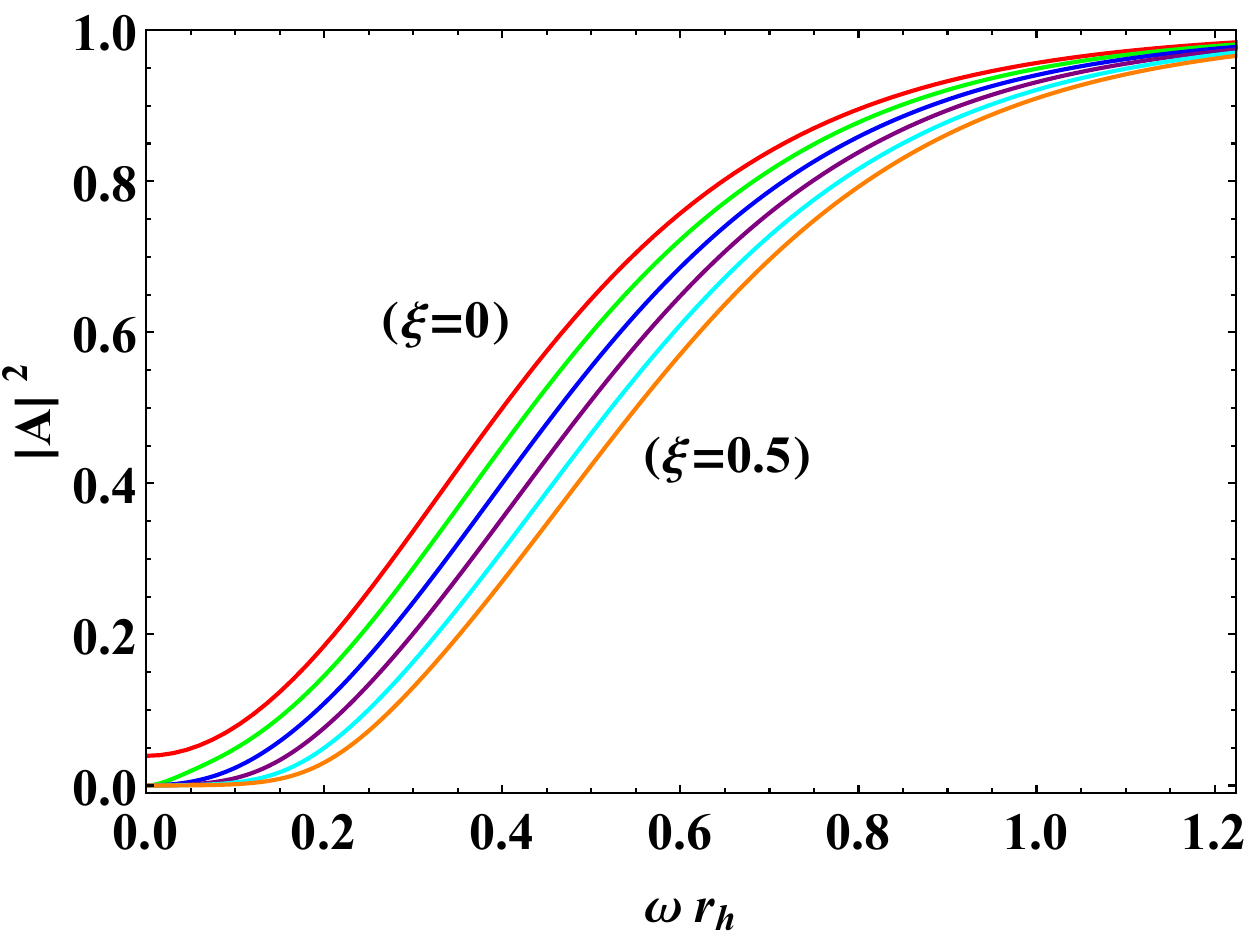}}
  \caption{(color online). Greybody factors for brane scalar fields for $l = 0$, $n = 2$,
$\Lambda = 0.1$ and variable (top to bottom) $\xi = 0, 0.1, 0.2, 0.3, 0.4, 0.5$.}
   \label{brane_xi}
  \end{center}
\end{figure}

We now turn to the dependence of the greybody factor on the two important parameters
of our model, the non-minimal coupling constant of the scalar field and the cosmological
constant of spacetime. In Fig. \ref{brane_xi}, we plot the dependence of the greybody factor
on $\xi$ for the dominant mode ($l = 0$) and for fixed cosmological constant
($\Lambda = 0.1$) and number of extra dimensions ($n = 2$). From the equation of
motion (\ref{field-eq-brane}) of the brane scalar field, we may interpret the coupling term
of the scalar field to the curvature scalar $R_4$ as an effective mass term for $\Phi$ --
this has been noted before in \cite{Crispino, KPP1}. As a result, the increase in the value
of $\xi$ corresponds to an increased effective mass, and thus to a suppressed transmission
probability for the particle, as found before in \cite{Page, Jung, Sampaio, KP1}. This is
in perfect agreement with the behaviour depicted in Fig. \ref{brane_xi}, where the greybody
factor gets suppressed the higher the value of $\xi$ becomes. Once again, for the minimal
coupling case of $\xi=0$, the non-vanishing low-energy asymptotic limit
(\ref{geom_brane}) is recovered as expected.

\begin{figure}[b!]
    \centering
    \begin{subfigure}[b]{0.318\textwidth}
        \includegraphics[width=\textwidth]{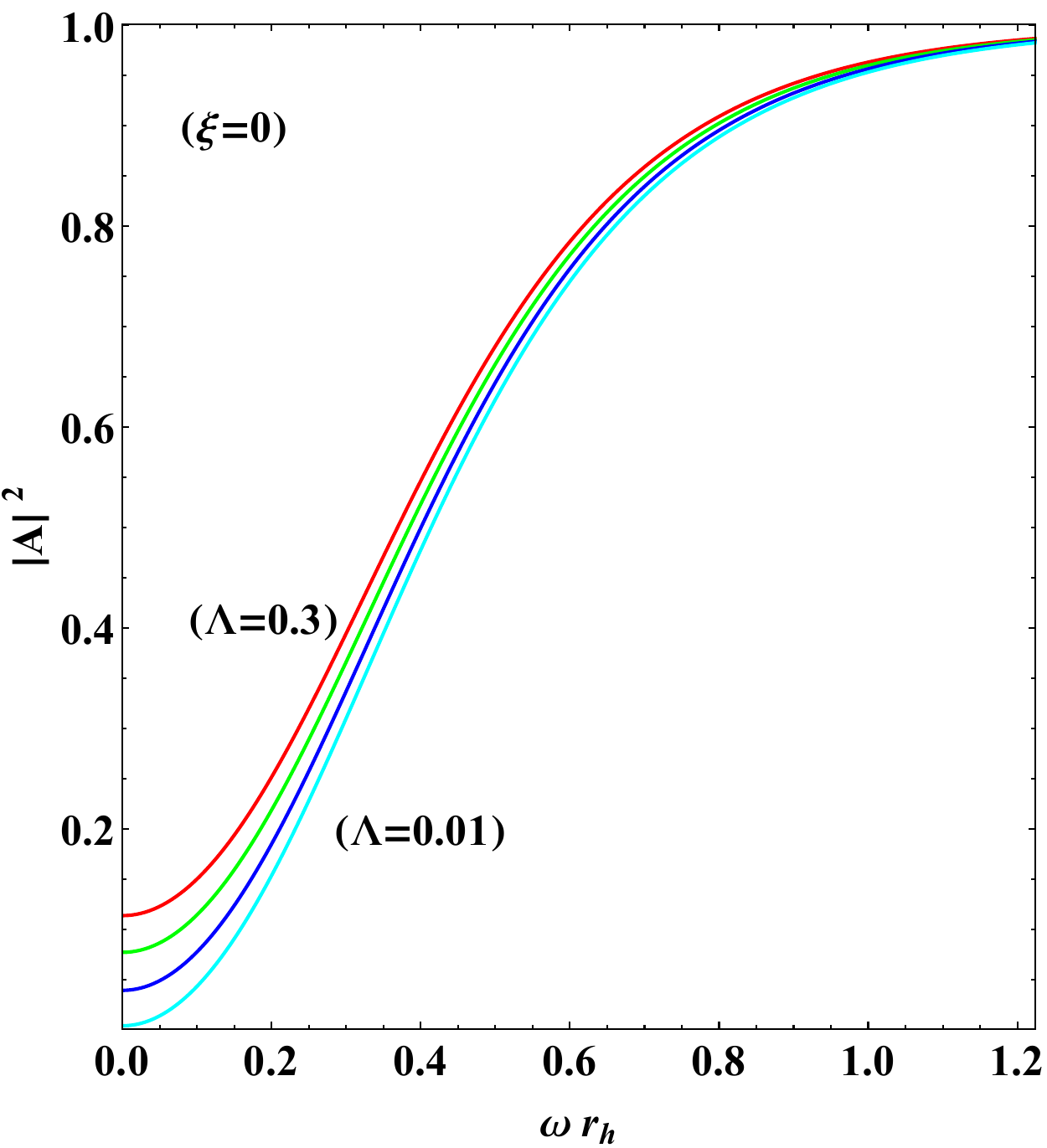}
        \caption{}
        \label{brane_xi00}
    \end{subfigure}
    ~ 
    \begin{subfigure}[b]{0.318\textwidth}
        \includegraphics[width=\textwidth]{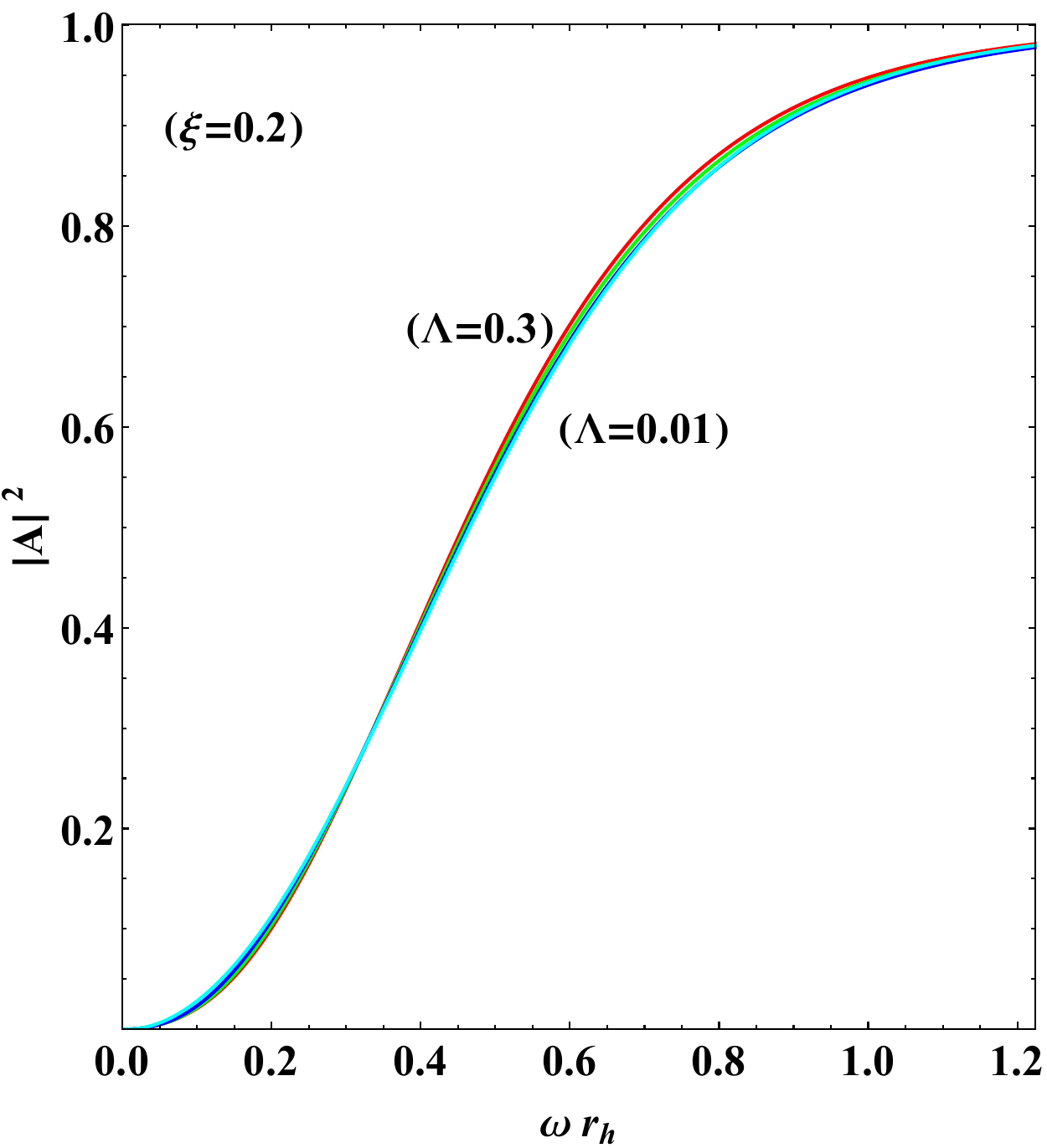}
        \caption{}
        \label{brane_xi02}
    \end{subfigure}
    ~ 
    \begin{subfigure}[b]{0.318\textwidth}
        \includegraphics[width=\textwidth]{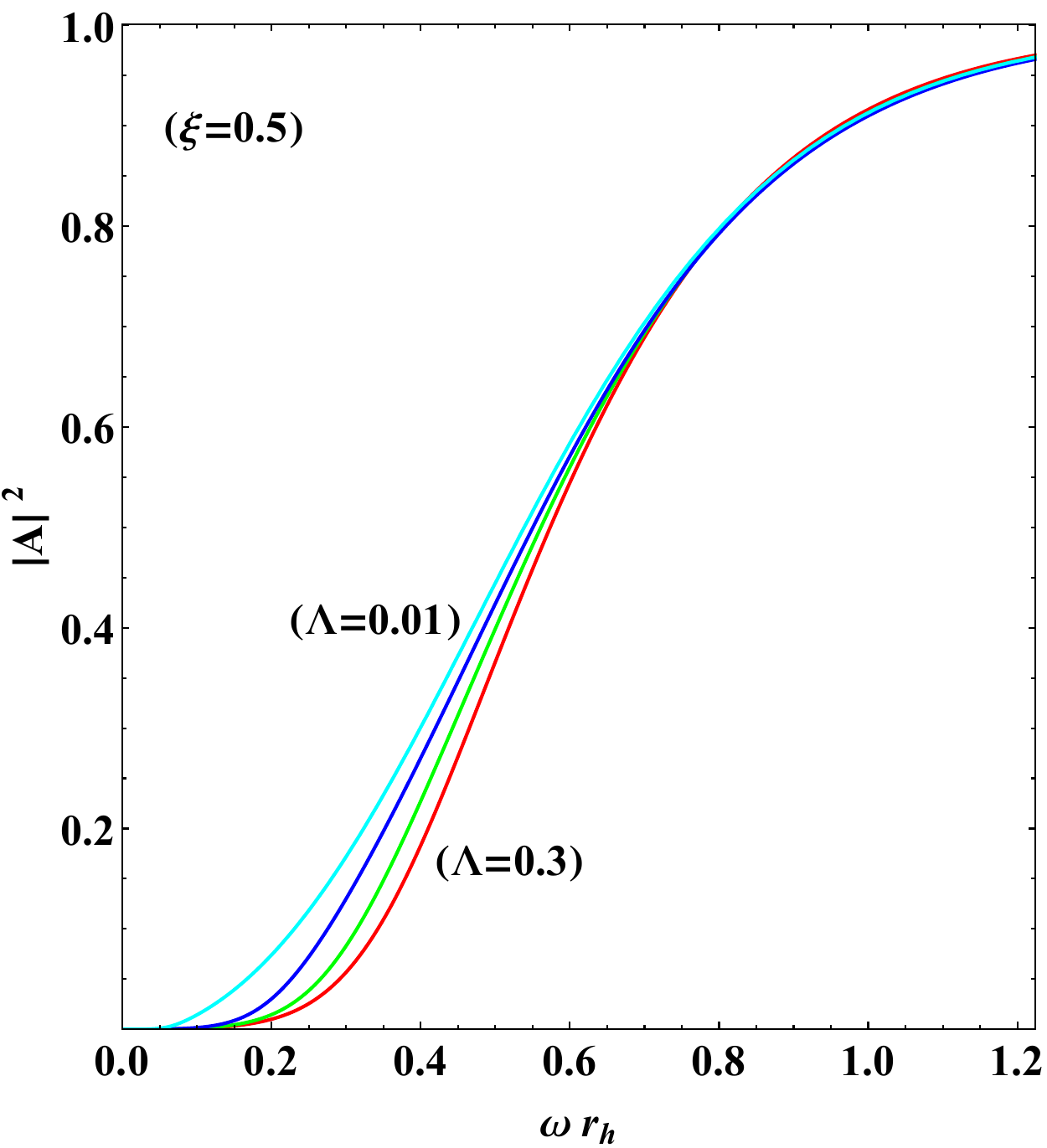}
        \caption{}
        \label{brane_xi05}
    \end{subfigure}
    \caption{(colour online). Greybody factors for brane scalar fields for $l = 0 , n = 2$ and
$\Lambda = 0.01,0.1,0.2,0.3$ and (a) for $\xi = 0$, (b) $\xi = 0.2$ and (c) $\xi=0.5$.
}\label{br_var_L}
\end{figure}

The effect of the cosmological constant on the greybody factor, as was discussed in \cite{KPP1},
depends on the value of the coupling constant $\xi$. In Fig. \ref{br_var_L}, we plot $|A|^2$ for
four different values of $\Lambda$ and three different values of $\xi$. For the minimal-coupling
case, depicted in Fig. \ref{br_var_L}(a), larger values of the cosmological constant enhance the
greybody factor throughout the energy regime. For the intermediate value $\xi = 0.2$
[see Fig. \ref{br_var_L}(b)], the greybody factors exhibit a slight interchange of behaviour at
the intermediate-energy regime but an extremely soft dependence on $\Lambda$ altogether.
As the coupling constant $\xi$ increases further, we see from Fig. \ref{br_var_L}(b) that
the role of $\Lambda$ is now
reversed, and larger values of the cosmological constant lead to a suppression in the 
greybody factor especially in the low-energy regime. As noted in \cite{KPP1}, this
subtle behaviour is due to the two different contributions of $\Lambda$: for small
values of $\xi$, the suppression due to the mass term (that is proportional also to $\Lambda$)
is small and the cosmological constant, as part of the metric function, subsidizes the
transmission probability of the brane scalar field through the potential barrier (\ref{pot_brane})
\cite{KGB}; for intermediate values of $\xi$, the two contributions almost cancel each other, while
for large values of $\xi$, the effective mass increases substantially leading to the suppression
of $|A|^2$ in terms of $\Lambda$.

%

\subsection{Transmission of Scalar Particles in the Bulk}

Scalar particles that propagate in the higher-dimensional gravitational background
(\ref{bhmetric}) obey the equation of motion (\ref{field-eq-bulk}). The greybody factor
may be determined in this case by solving the radial equation (\ref{radial_bulk}).
This equation was also solved analytically in \cite{KPP1} by following a similar
approximate method. In fact, the asymptotic solutions near the black-hole
and cosmological horizons take the exact same forms as (\ref{BH-gen}) and (\ref{FF2-dS}),
respectively, differing only in the definition of the hypergeometric indices and in the
power coefficients \cite{KPP1} --
in fact the power coefficient $\alpha_1$ adopts exactly the same functional form
(\ref{sol-albe-rh}) as in the brane emission. 

Apart from the above modifications, the procedure that was followed was the same as for
propagation on the brane. The matching of the two asymptotic solutions led again to an
analytic expression for the ratio $B_2/B_1$ and, through Eq. (\ref{greybody}), for the greybody
factor for bulk scalar fields. This expression was shown again to correctly reproduce, for the
minimal-coupling case, the non-vanishing, geometric, low-energy limit for $|A|^2$ for the
mode $l=0$ found in \cite{KGB} and now given by
\beq
|A^2|=
\frac{4 (r_hr_c)^{(n+2)}}{(r_c^{n+2}+r_h^{n+2})^2}+ {\cal O}(\omega)\,.
\label{geom_bulk}
\eeq
Comparing the above expression with the one on the brane (\ref{geom_brane}), we observe that
the presence of the number of extra dimensions $n$ in the exponents of $r_h$ and $r_c$ causes
the magnitude of the low-energy asymptotic value in the bulk to be significantly smaller
compared to the one on the brane. For $\xi \neq 0$, the greybody factors for all modes
of the bulk scalar field reduce again to zero \cite{KPP1}.

Once again, the smooth matching of the asymptotic solutions was achieved under the assumptions
of small cosmological constant and small coupling constant. As a result, we turn again to the
numerical integration of the radial equation (\ref{radial_bulk}) in order to produce exact, complete
results for the greybody factor. Given that the asymptotic solutions near the black-hole and
cosmological horizons are the same for brane and bulk propagation, the same holds true
for their expanded forms (\ref{BH-exp}) and (\ref{CO-exp}) and the boundary conditions
for the numerical integration (\ref{R_BH_num})-(\ref{dR_BH_num}). Once again, the integration
starts from the proximity of the
black-hole horizon and proceeds to the cosmological horizon where the amplitudes of the
incoming and outgoing modes are isolated and determined through
Eqs. (\ref{incoming})-(\ref{outgoing}). The exact value of the greybody factor, for arbitrary
values of the particle and spacetime parameters, is then found via Eq. (\ref{greybody}).

\begin{figure}[t!]
    \centering
    \begin{subfigure}[ht]{0.47\textwidth}
        \includegraphics[width=\textwidth]{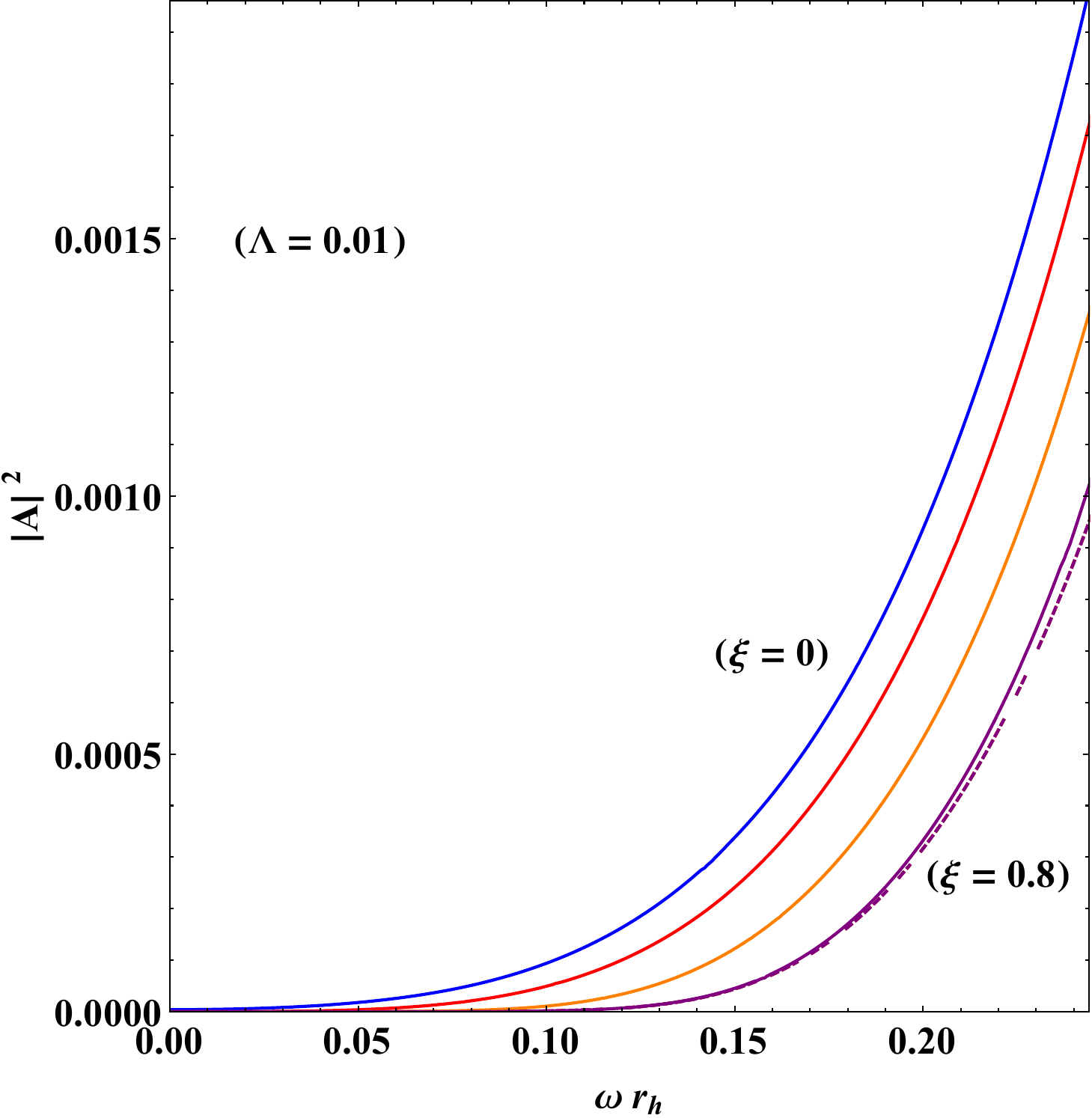}
        \caption{}
    \end{subfigure}
    ~ 
    \begin{subfigure}[ht]{0.48\textwidth}
        \includegraphics[width=\textwidth]{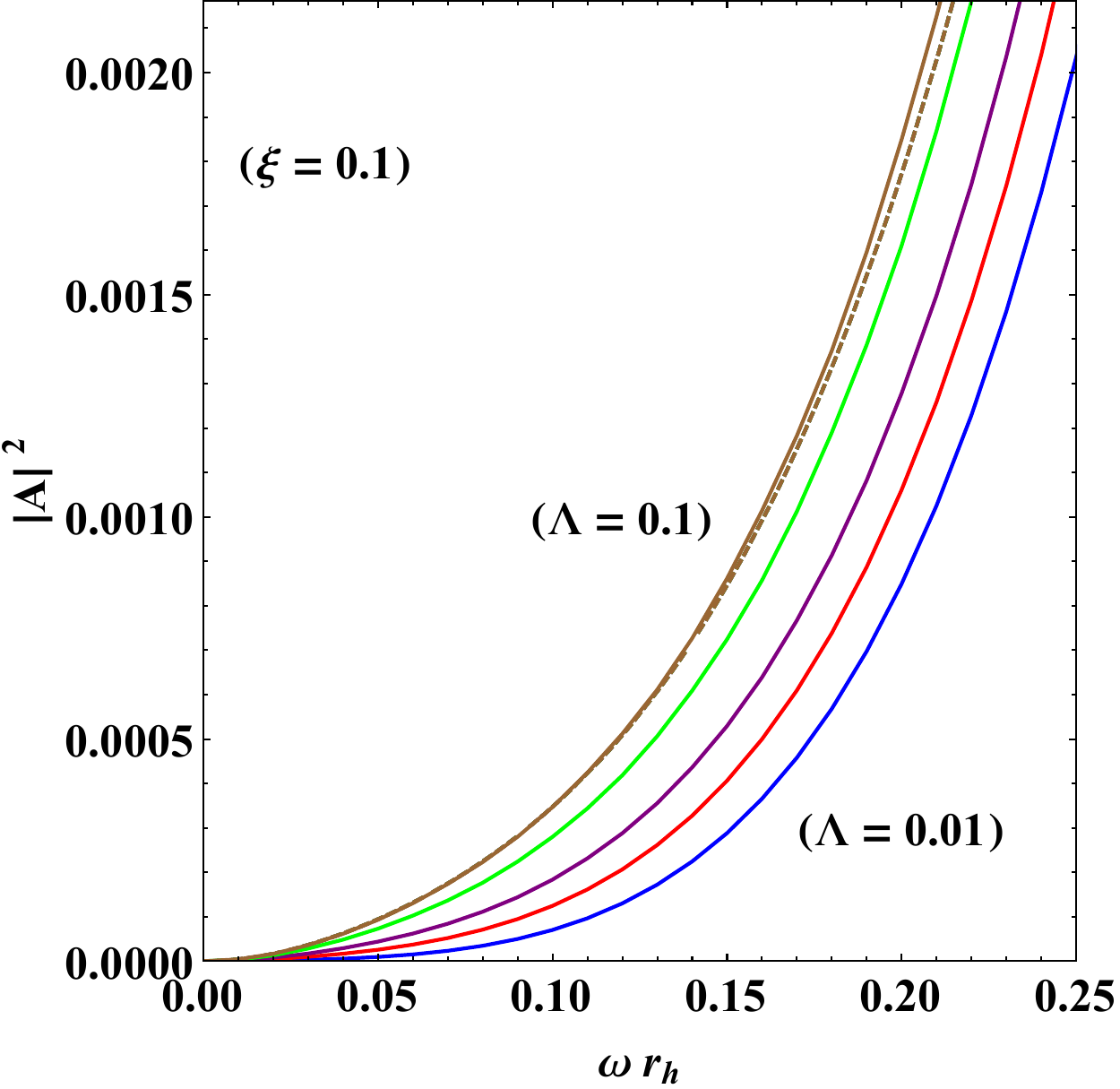}
        \caption{}
    \end{subfigure}
    \caption{(colour online). Greybody factors for bulk scalar fields. Analytical (dashed curves)
and numerical (solid curves) results for $l = 0 , n = 2$ and: (a) for $\Lambda = 0.01$ and
variable $\xi = 0, 0.2, 0.5, 0.8$; (b) for $\xi = 0.1$ and variable
$\Lambda = 0.01, 0.03, 0.05, 0.08, 0.1$.   }\label{anal_vs_num_bu_xi_L}
\end{figure}

For completeness, the comparison of analytical and numerical results for $|A|^2$ for scalar
fields in the bulk is presented in Figs. \ref{anal_vs_num_bu_xi_L}(a,b) in terms of the coupling
constant $\xi$ and cosmological constant $\Lambda$, respectively. An excellent agreement
is observed between the exact (dashed) and the analytical (solid) results for the greybody
factors that persists beyond the low-energy regime. As the values of either $\xi$ or $\Lambda$
become larger, deviations appear, as expected, and the validity regime of the analytic
results becomes smaller.

\begin{figure}[t!]
    \centering
    \begin{subfigure}[ht]{0.465\textwidth}
        \includegraphics[width=\textwidth]{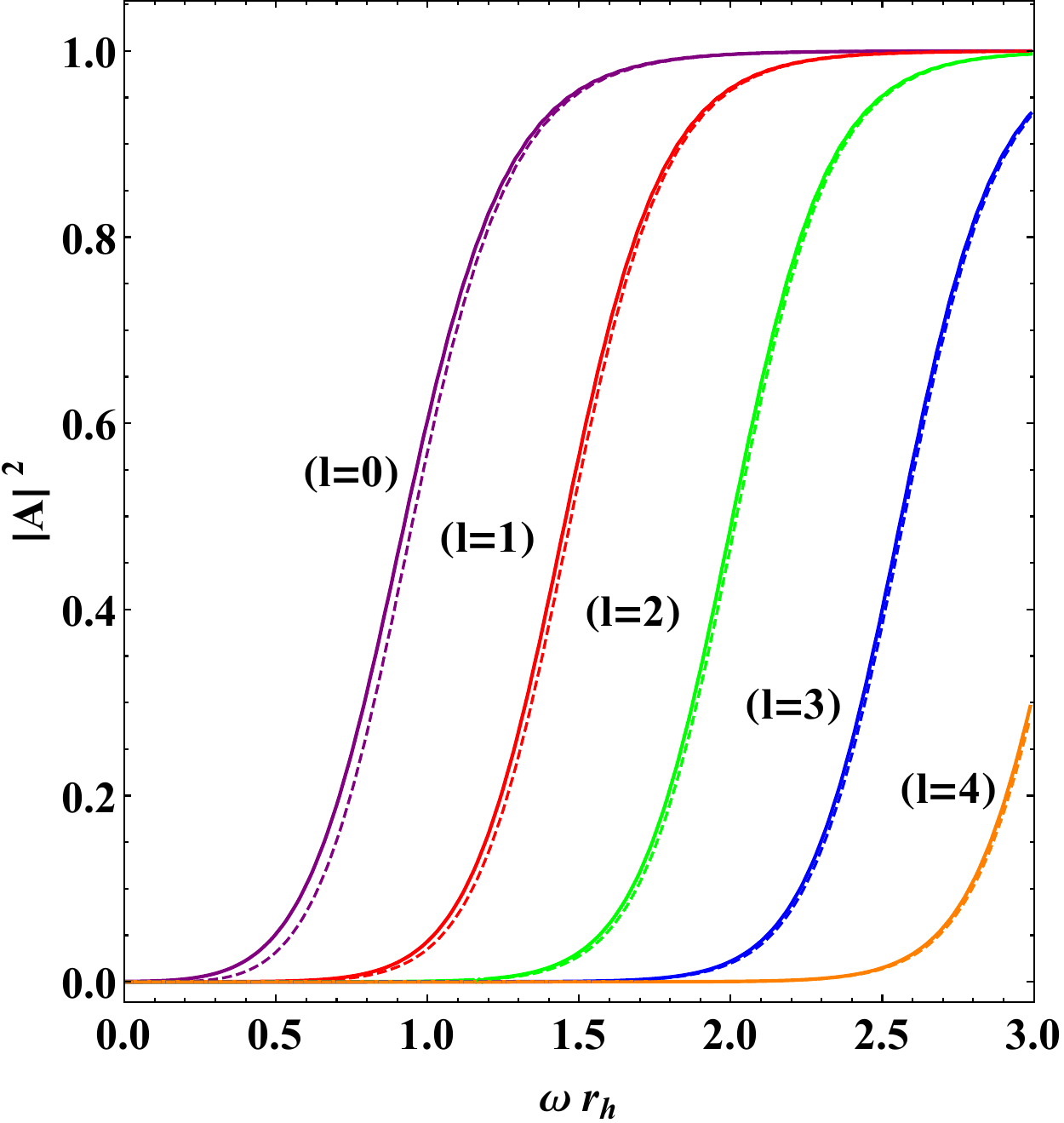}
        \caption{}
    \end{subfigure}
    ~ 
    \begin{subfigure}[ht]{0.45\textwidth}
        \includegraphics[width=\textwidth]{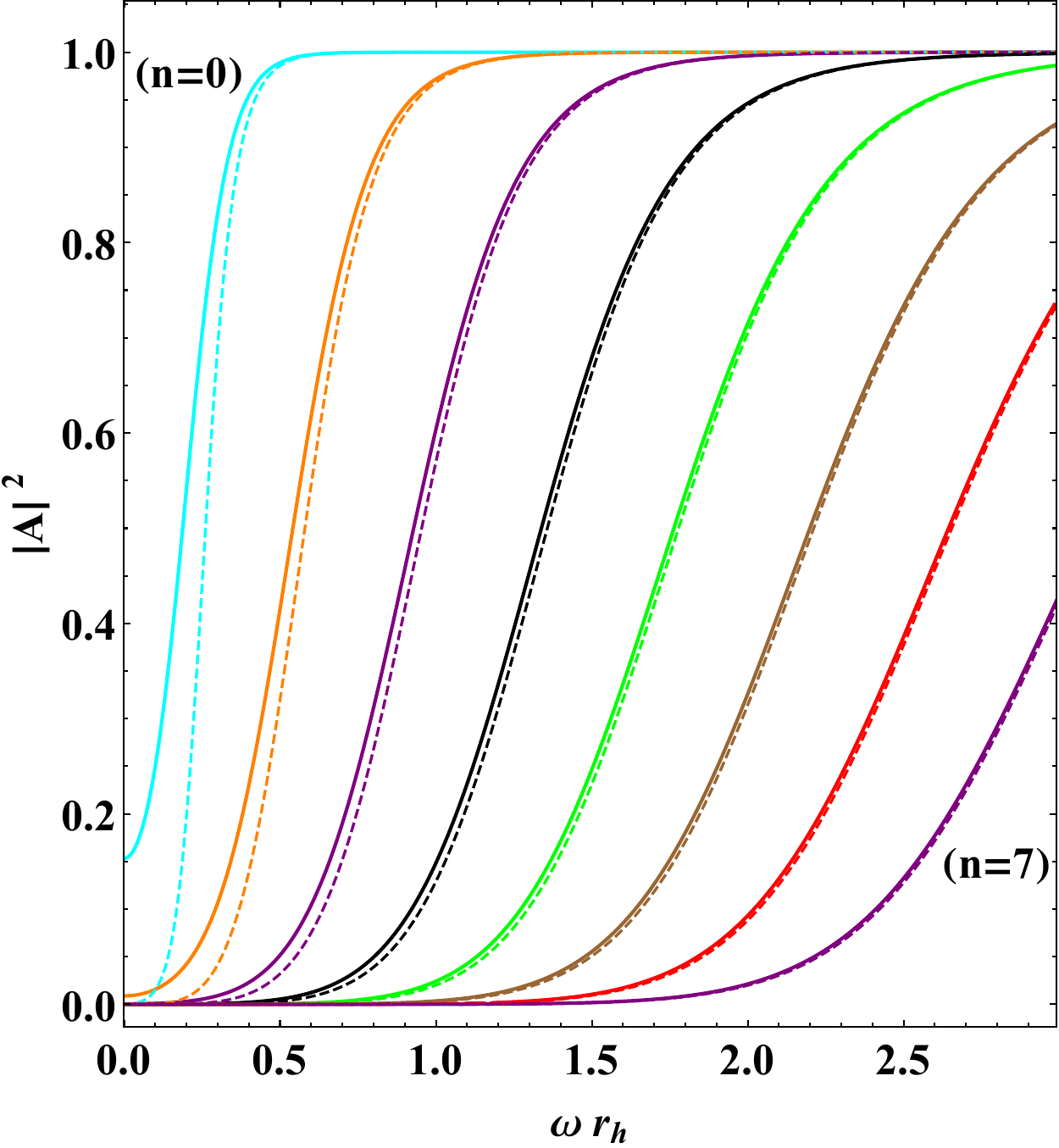}
        \caption{}
    \end{subfigure}
    \caption{(color online). Greybody factors for bulk scalar fields for $\Lambda = 0.1$,
$\xi = 0$ (solid curves) or $\xi = 0.3$ (dashed curves) and: (a) $n = 2$ and variable
$l = 0,1,2,3,4$; (b) $l = 0$ and variable $n = 0, 1, 2, 3, 4, 5, 6, 7$. } \label{bu_l_and_n}
\end{figure}

Turning now to the complete, exact results, in Fig. \ref{bu_l_and_n}(a) we depict the 
dependence of the greybody factor for scalar fields in the bulk in terms of the angular
momentum number $l$, for $n=2$ and $\Lambda=0.1$ (in units again of $r_h^{-2}$).
The suppression of $|A|^2$ as the partial mode number increases
is observed for both cases of minimal (solid curves) and non-minimal (dashed curves)
coupling; however, the difference is less noticeable than the one observed in the case of
brane propagation, even for small values of $l$. Our exact numerical analysis
has led to the complete greybody curves in contrast to our previous analytic study where
these curves were restricted to the very low-energy regime due to the existence of poles.
Because of the smallness of the asymptotic geometric value of $|A|^2$ for the lowest mode
$l=0$ in the limit $\omega \rightarrow 0$ for the case $n=2$, the difference between
the minimal and non-minimal coupling cases can not be discerned in Fig. \ref{bu_l_and_n}(a)
-- a zoom-in plot of the low-energy regime would be necessary to achieve this. 

This difference is however visible in Fig. \ref{bu_l_and_n}(b), where we show the dependence
of the greybody factor on $n$ for the dominant mode with $l=0$:
for the two lowest values, $n=0$ and $n=1$, the difference in the low-energy asymptotic
values of $|A|^2$, as we turn on or off the coupling constant $\xi$, may be clearly seen.
In general, the greybody factor is universally suppressed, as the number of extra spacelike
dimensions increases, and the scalar particle is less likely to overcome the potential barrier.
Note that our exact numerical analysis has provided us with results for all values of $n$,
in contrast to our analytic approach where results for only even values of $n$ were derived
due to the existence again of poles. 

\begin{figure}[t!]
  \begin{center}
\mbox{\includegraphics[width = 0.52 \textwidth] {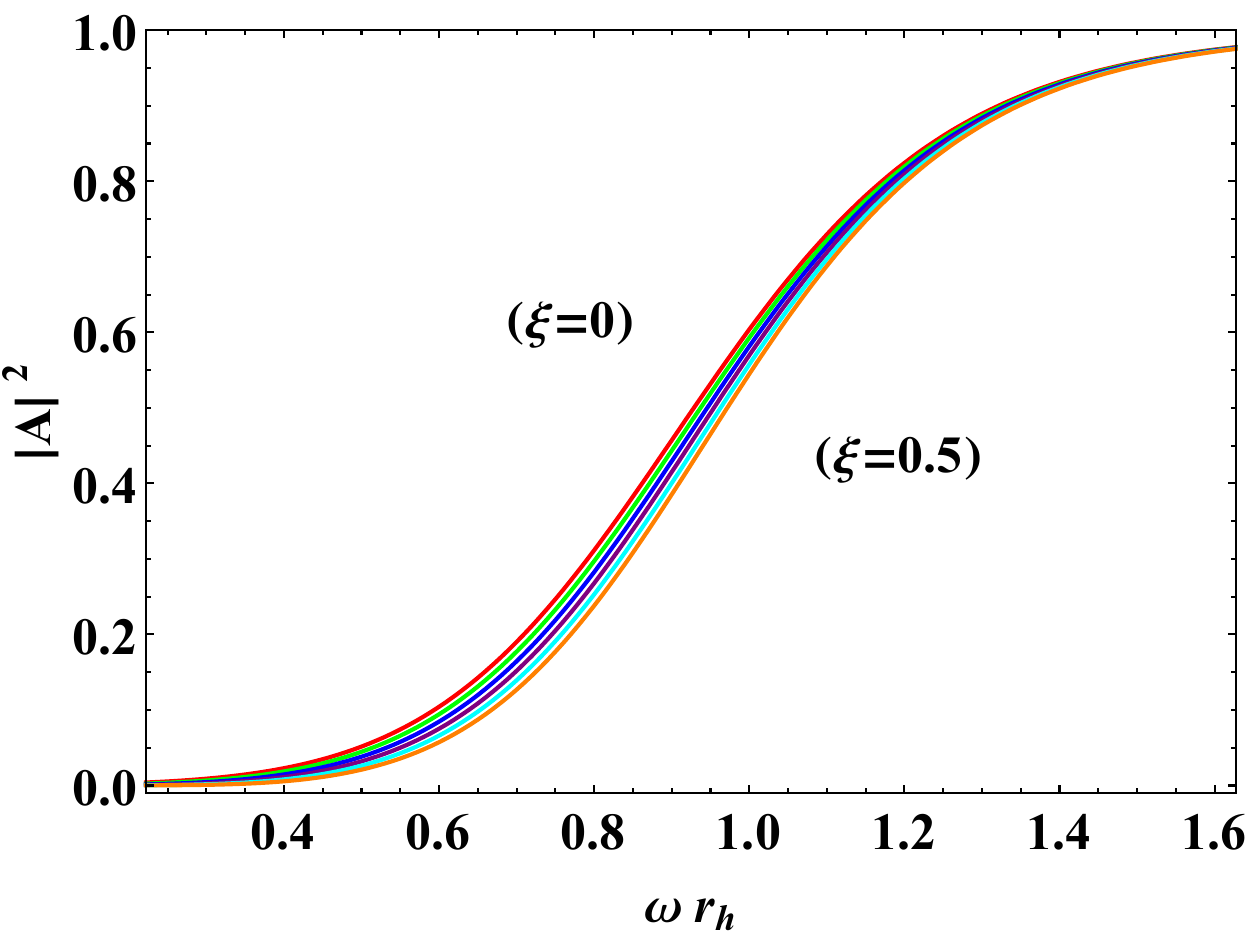}}
  \caption{(color online). Greybody factors for bulk scalar fields for $n = 2 , \Lambda = 0.1 , l = 0$ and variable $\xi = 0, 0.1, 0.2, 0.3, 0.4, 0.5$.}
   \label{bulk_xi}
  \end{center}
\end{figure}

In Fig. \ref{bulk_xi}, the dependence of the greybody factor on the coupling constant $\xi$ is
depicted for the dominant mode ($l=0$). We observe that, in the bulk, in contrast to the brane
emission, for the same choice of parameters, the effect of the variation in $\xi$ is milder --
this is justified by the fact that the dependence of the bulk potential barrier on $\xi$ is also
very mild as was shown in \cite{KPP1}. Still, as the coupling constant increases, the greybody factor
gets suppressed; this behavior can be once again interpreted as the result of the increased
effective mass of the field generated through its coupling to the Ricci scalar.

\begin{figure}[b!]
    \centering
    \begin{subfigure}[ht]{0.46\textwidth}
        \includegraphics[width=\textwidth]{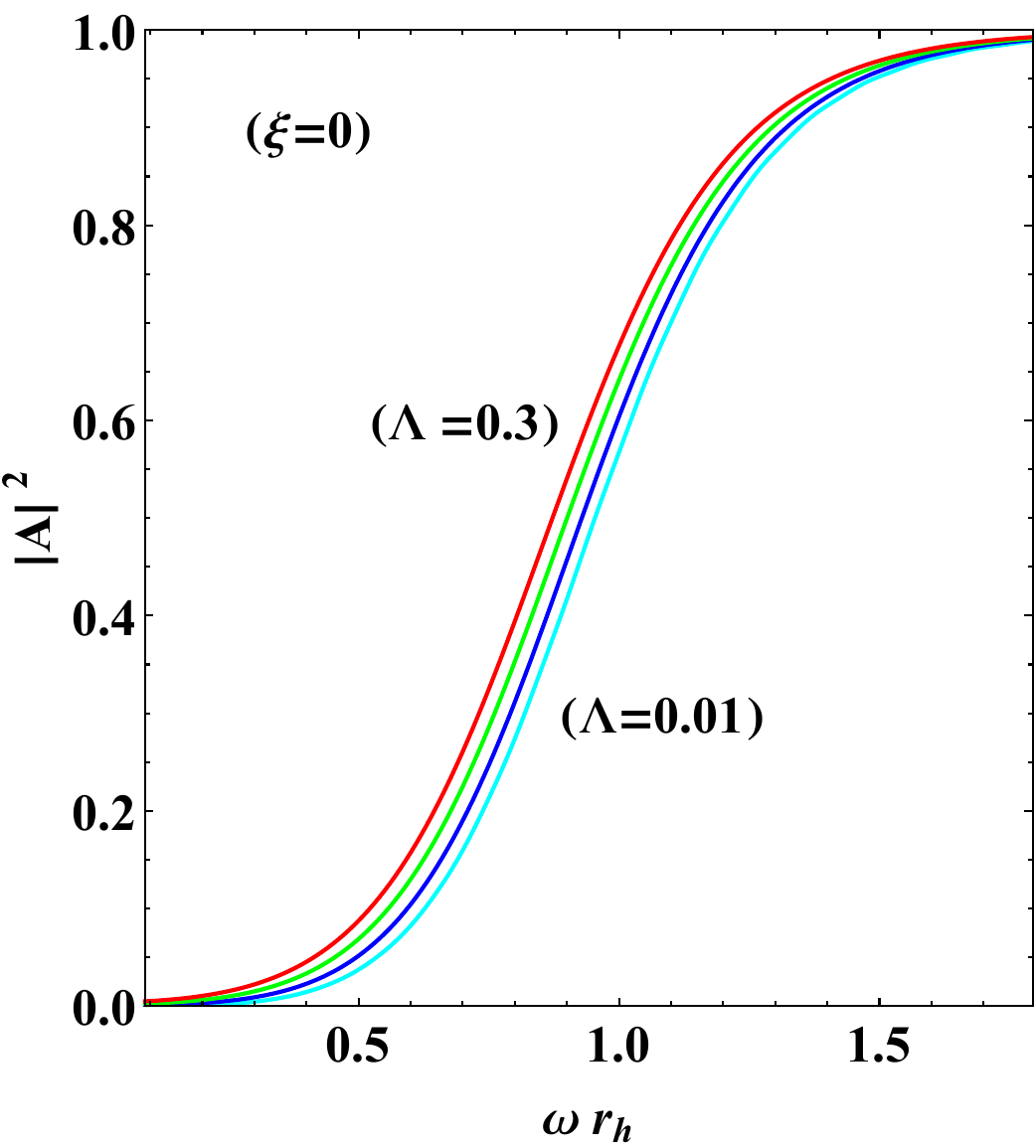}
        \caption{}
        \label{bulk_xi00}
    \end{subfigure}
    ~ 
    \begin{subfigure}[ht]{0.46\textwidth}
        \includegraphics[width=\textwidth]{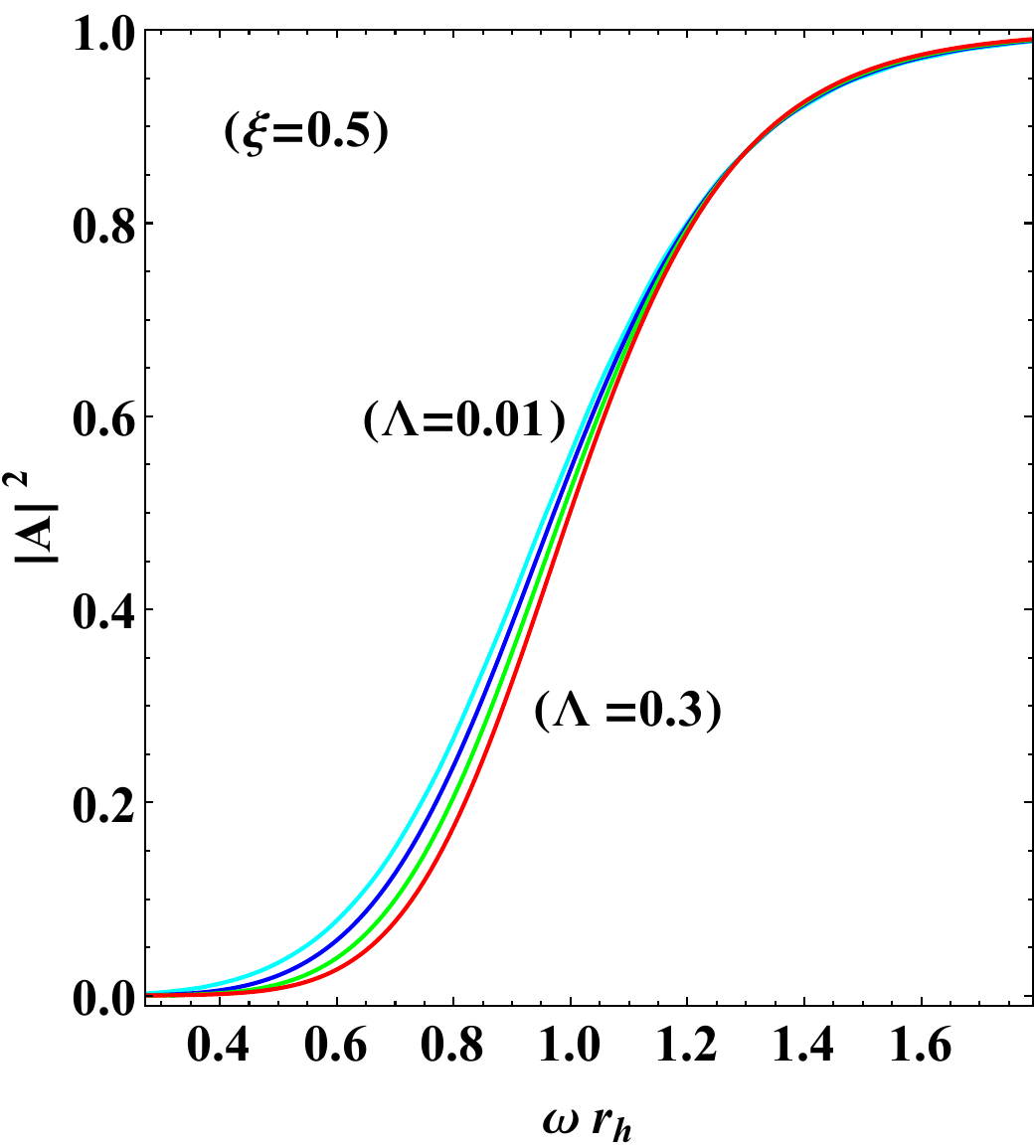}
        \caption{}
        \label{bulk_xi05}
    \end{subfigure}
    \caption{(colour online). Greybody factors for bulk scalar fields for $l = 0 , n = 2$ and $\Lambda = 0.01,0.1,0.2,0.3$ and (a) for $\xi = 0$ and (b) $\xi = 0.5$.  }\label{bu_Lambda}
\end{figure}

The effect of the cosmological constant on the greybody factor for bulk scalar fields is shown in
Fig. \ref{bu_Lambda}: the minimal coupling case is given in Fig. \ref{bu_Lambda}(a) while a
non-vanishing value of the field coupling ($\xi = 0.5$) is employed in Fig. \ref{bu_Lambda}(b).
The dual role of the cosmological constant -- contributing simultaneously to the lowering of
the potential barrier and to the effective field mass -- is again clear. Although the dependence
of $|A|^2$ on $\Lambda$ is in general soft, for vanishing or small values of the coupling constant
$\xi$, the greybody factor is enhanced with the cosmological constant; for larger values of $\xi$
though, the situation is reversed, mainly in the low and intermediate-energy regime, where an increase
in $\Lambda$ suppresses the greybody factor. It is worth noting that a very good agreement is 
found between the numerical results presented here and the analytical ones derived in \cite{KPP1}
regarding the role of the cosmological constant. 

Finally, let us finish this section by making the following observations: by mere comparison of
Figs. 2(a,b) and 6(a,b) depicting the dependence of the greybody factors for brane and bulk
scalar fields, respectively, on $l$ and $n$ -- a comparison made easy by our selecting
on purpose the same sets of parameters -- one may see that the suppression of $|A|^2$
with both $l$ and $n$ is much more important for bulk rather than for brane propagation.
This effect is a well-known one leading to the dominance of the brane over the bulk channel when
one studies the emission of scalar fields by a higher-dimensional Schwarzschild black hole
\cite{KMR, HK1}. However, in the present analysis we have two more parameters, the
coupling constant of the scalar field and the cosmological constant of the spacetime
background. Regarding the former, we may again observe, from Figs. 3 and 7, that the
suppression it causes to the value of the greybody factor is more prominent for
brane scalars than for bulk scalars. Will the effect of the non-minimal
coupling constant be able to undermine the dominance of the brane emission channel?
And, will the cosmological constant with its subtle effect, despite the soft dependence of the
greybody factor on it, be able to affect in any way the bulk-to-brane energy balance?
We will return to those questions at the final part of the following section.


\section{Energy Emission Rates}

We now proceed to the derivation of the differential energy emission rates by a higher-dimensional
Schwarzschild-de-Sitter black hole in the form of scalar fields. We will study the emission of
scalar Hawking radiation both
on the brane and in the bulk and, subsequently, discuss the total emissivity of each channel
and their relative ratio in order to investigate whether the additional parameters of the model, the
coupling constant and the cosmological constant, may change the energy balance between
brane and bulk.

\subsection{Power Spectra for Emission on the Brane}

We start with the emission of scalar particles on the brane. The differential energy emission
rate is given by the expression \cite{HK1, Kanti:2004, KGB}
\beq
\frac{d^2E}{dt\,d\omega}=\frac{1}{2\pi}\,\sum_l\,\frac{N_l\,|A|^2\,\omega}{\exp(\omega/T_{BH})-1}\,,
\label{diff-rate-brane}
\eeq
where $\omega$ is the energy of the emitted, massless particle, $|A|^2$ the greybody factor,
or transmission probability, computed in the previous section, and $N_l=2l+1$ the multiplicity of
states that, due to the spherical symmetry, have the same angular-momentum number.
Also, $T_{BH}$ is the temperature of the black hole determined through the surface gravity as
\beq
T_{BH}=\frac{k_H}{2\pi}=\frac{1}{4\pi r_h}\,\Bigl[(n+1)-\frac{2 \kappa^2_D \Lambda r_h^2}
{(n+2)}\Bigr]\,,
\label{T_BH_0}
\eeq
where Eq. (\ref{mu}) has been used to eliminate any dependence on the mass parameter $\mu$.
However, here we will follow the Bousso-Hawking definition of the temperature since Eq. (\ref{T_BH_0})
is strictly valid only when the spacetime is asymptotically flat \cite{Bousso}. For the
Schwarzschild-de-Sitter spacetime, though, this does not hold; the only point where ``asymptotic flatness'' may be
considered to hold
is at the point $r_0$ located between $r_h$ and $r_c$ where the effects of black-hole attraction
and cosmological repulsion cancel out. This point corresponds to an extremum of the metric
function $h(r)$ and is found to be
\beq
r_0=\Bigl[\frac{(n+1)(n+2) (n+3) \mu}{4 k_D^2 \Lambda}\Bigr]^{1/(n+3)}\,.
\label{r0-def}
\eeq
Therefore, the correct definition of the temperature of a Schwarzschild-de-Sitter black hole is
the following  \cite{KGB, Bousso}
\beq
T_{BH}=\frac{k_H}{2\pi}=\frac{1}{\sqrt{h(r_0)}}\,\frac{1}{4\pi r_h}\,\Bigl[(n+1)-
\frac{2 \kappa^2_D \Lambda r_h^2}{(n+2)}\Bigr]\,.
\label{T_BH}
\eeq
Note that, as expected, when the cosmological constant tends to zero, the point $r_0$ moves to
infinity: there, the metric function becomes unity, the normalising factor $1/\sqrt{h(r_0)}$
disappears and Eq. (\ref{T_BH}) gives the well-known expression for the temperature of a 
higher-dimensional Schwarzschild black hole~\footnote{The temperature of the universe $T_C$
is also given by an expression similar to Eq. (\ref{T_BH_0}) with $r_h$ being replaced by
$r_c$ and an overall minus sign introduced to ensure the positivity of $T_C$. The latter may
be alternatively written as $T_C=-A_c/4\pi r_c$, therefore the quantity $A_c$ is negative as used
in Section 3.1. Finally, one may easily conclude that $T_{BH}>T_C$, therefore it is only the
emission from the black-hole horizon that is relevant here.}.

\begin{figure}[t]
  \begin{center}
\mbox{\includegraphics[width = 0.46 \textwidth] {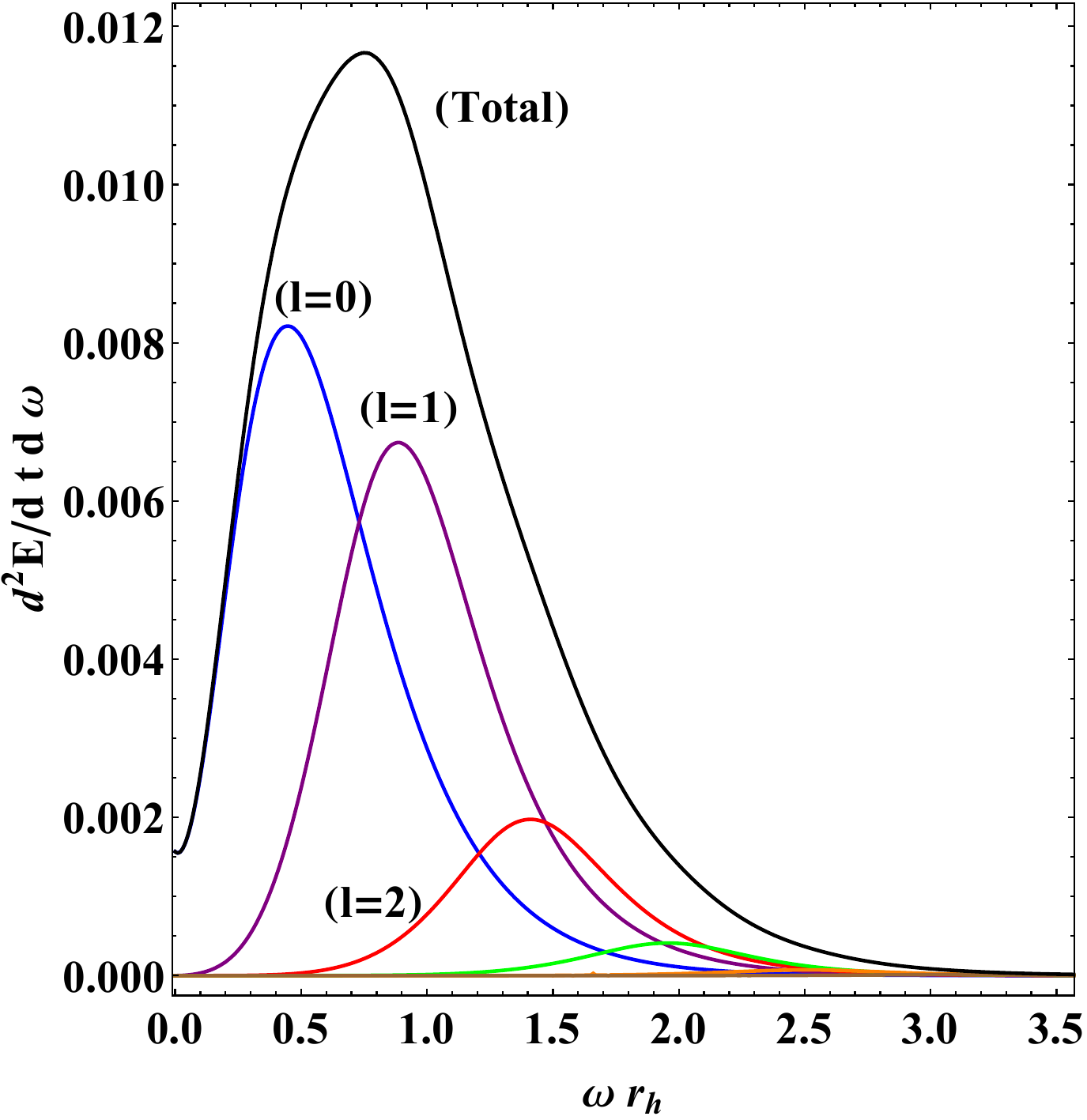}}
  \caption{(color online). Energy emission rate curves for brane scalar fields for $n = 2$,
$\Lambda = 0.1$, $\xi = 0$ for the first five dominant modes with $l = 0, 1, 2, 3, 4, 5$.}
   \label{eer_br_modes}
  \end{center}
\end{figure}

We now have all the tools to compute the energy emission rate on the brane: using
Eqs. (\ref{diff-rate-brane}) and (\ref{T_BH}) and the exact numerical results for the
greybody factor derived in Section 3.1, we may investigate the dependence of the spectrum on
all the particle and spacetime properties. The contribution to the energy emission rate comes
mainly from the dominant modes of the scalar field namely the ones with the lowest values
of $l$. As $l$ increases, the $l$-th mode contributes less and less to the total spectrum.
We took advantage of this fact to terminate the infinite sum in Eq. (\ref{diff-rate-brane})
at a finite number of modes. More precisely, we found that all modes higher than the $l = 7$
mode contribute an amount which is many orders of magnitude lower than the peak of the
power curve, and so we may safely ignore them. An indicative example of this behaviour is
shown in Fig. \ref{eer_br_modes} for $n=2$, $\Lambda=0.1$ and $\xi=0$: clearly, as $l$
increases, the modes contribute less to the total sum with the $l=5$ mode being already
irrelevant.

\begin{figure}[b!]
    \centering
    \begin{subfigure}[ht]{0.45\textwidth}
        \includegraphics[width=\textwidth]{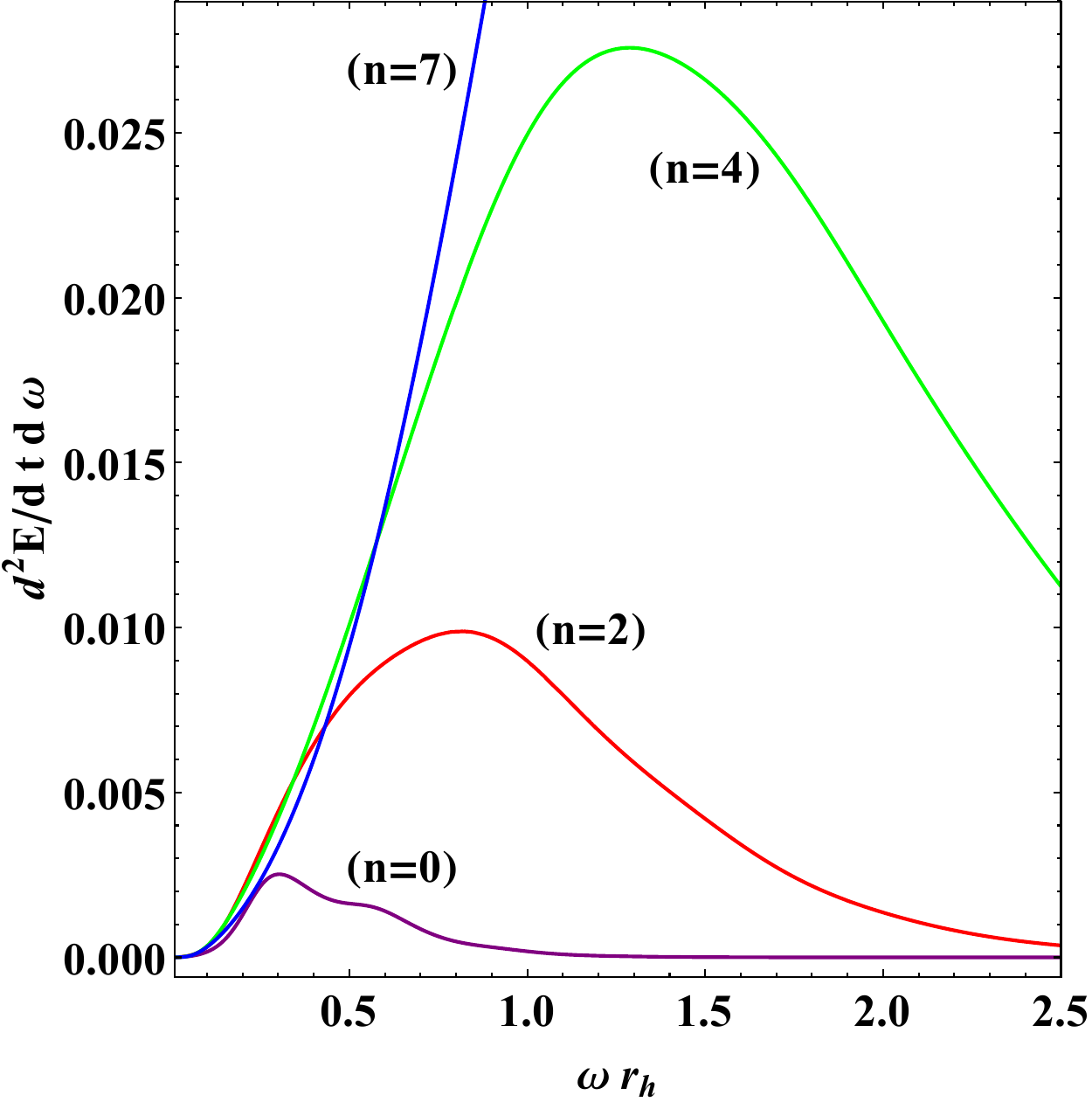}
        \caption{}
        \label{eer_br_n}
    \end{subfigure}
    ~ 
    \begin{subfigure}[ht]{0.45\textwidth}
        \includegraphics[width=\textwidth]{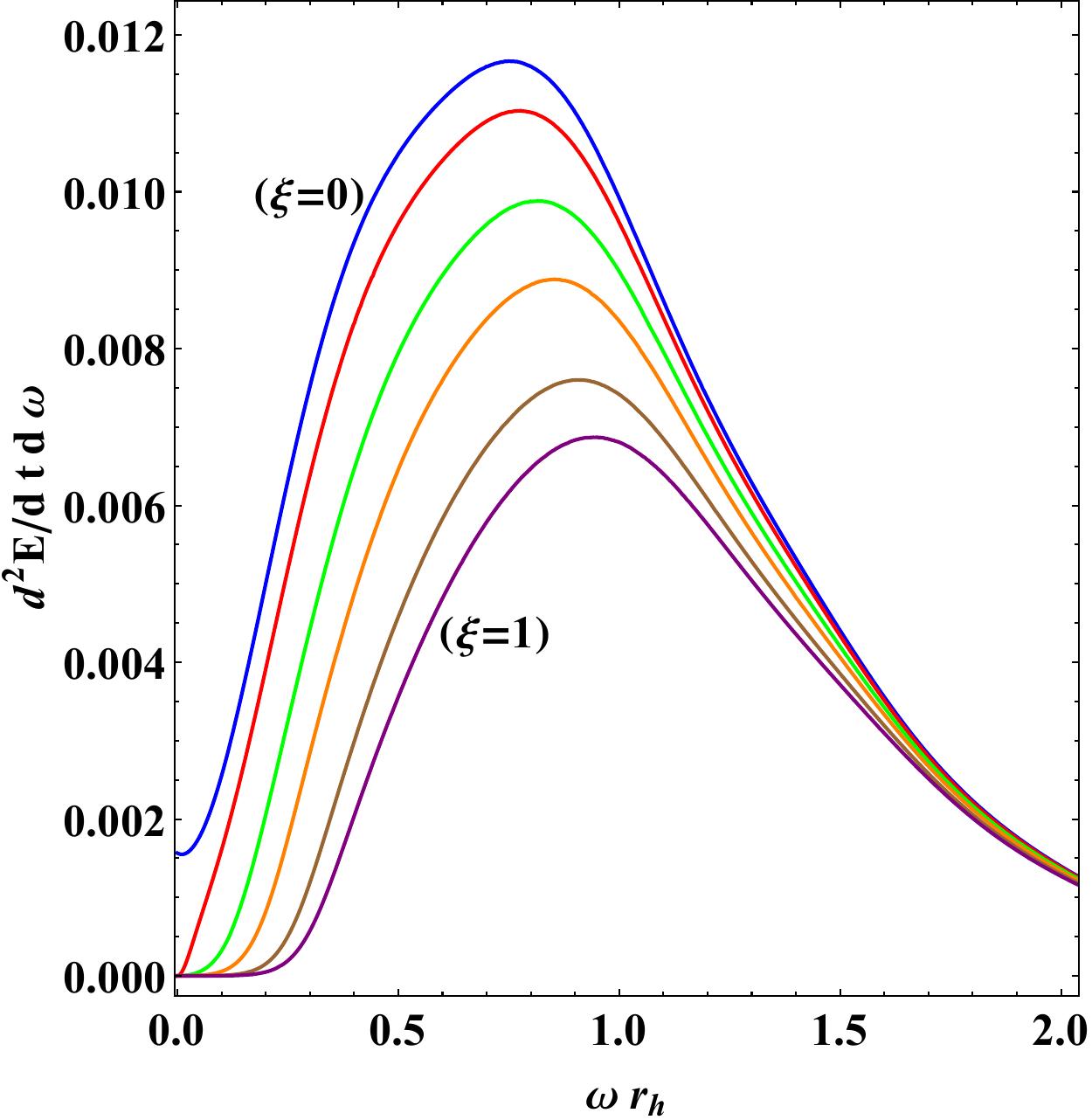}
        \caption{}
        \label{eer_br_xi}
    \end{subfigure}
    \caption{(colour online). Energy emission rates for brane scalar fields, for $\Lambda = 0.1$,
and: (a) $\xi = 0.3$ and variable $n = 0, 2, 4, 7$; (b) $n = 2$ and variable
$\xi = 0, 0.1, 0.3, 0.5, 0.8, 1$.   }\label{eer_br_n_xi}
\end{figure}

The dependence of the energy emission rate for brane scalars on the number of extra spacelike
dimensions is depicted in Fig. \ref{eer_br_n_xi}(a). Clearly, the energy emitted by the black hole
per unit time and unit frequency on the brane is enhanced with $n$; this feature was noted before
 in \cite{KGB} where the emission of a Schwarzschild-de-Sitter black hole was also studied as well
as in a number of previous works on both spherically and axially symmetric black holes
\cite{KMR}-\cite{CDKW3}. Although the greybody factor, as shown in section 3.1, gets suppressed
with the number of extra dimensions, the temperature of the black hole (\ref{T_BH_0}) gets
significantly enhanced leading at the end to the overall enhancement of the power spectra. 

On the other hand, as the temperature of the black hole is insensitive to the particle properties,
we expect the suppression of the greybody factor with the value of the non-minimal coupling
constant to carry on to the energy emission rate. From Fig.  \ref{eer_br_n_xi}(b), we observe
that this is indeed the case: as $\xi$ increases, the power spectra are suppressed throughout
the energy regime. This picture is also consistent with the interpretation of the non-minimal
coupling term of the scalar field to the Ricci curvature as an effective mass term: in \cite{KP1},
the power spectra for massive scalar fields were derived and the similarity between those
results and the behaviour depicted in Fig. \ref{eer_br_n_xi}(b) is striking; in both analyses,
the emission
curves adopt non-zero values at a gradually larger value of $\omega r_h$ as the mass
increases, exhibit a uniform suppression over the whole energy regime and converge to
a common ``tail'' at the high-energy regime.

\begin{figure}[t!]
    \centering
    \begin{subfigure}[b]{0.318\textwidth}
        \includegraphics[width=\textwidth]{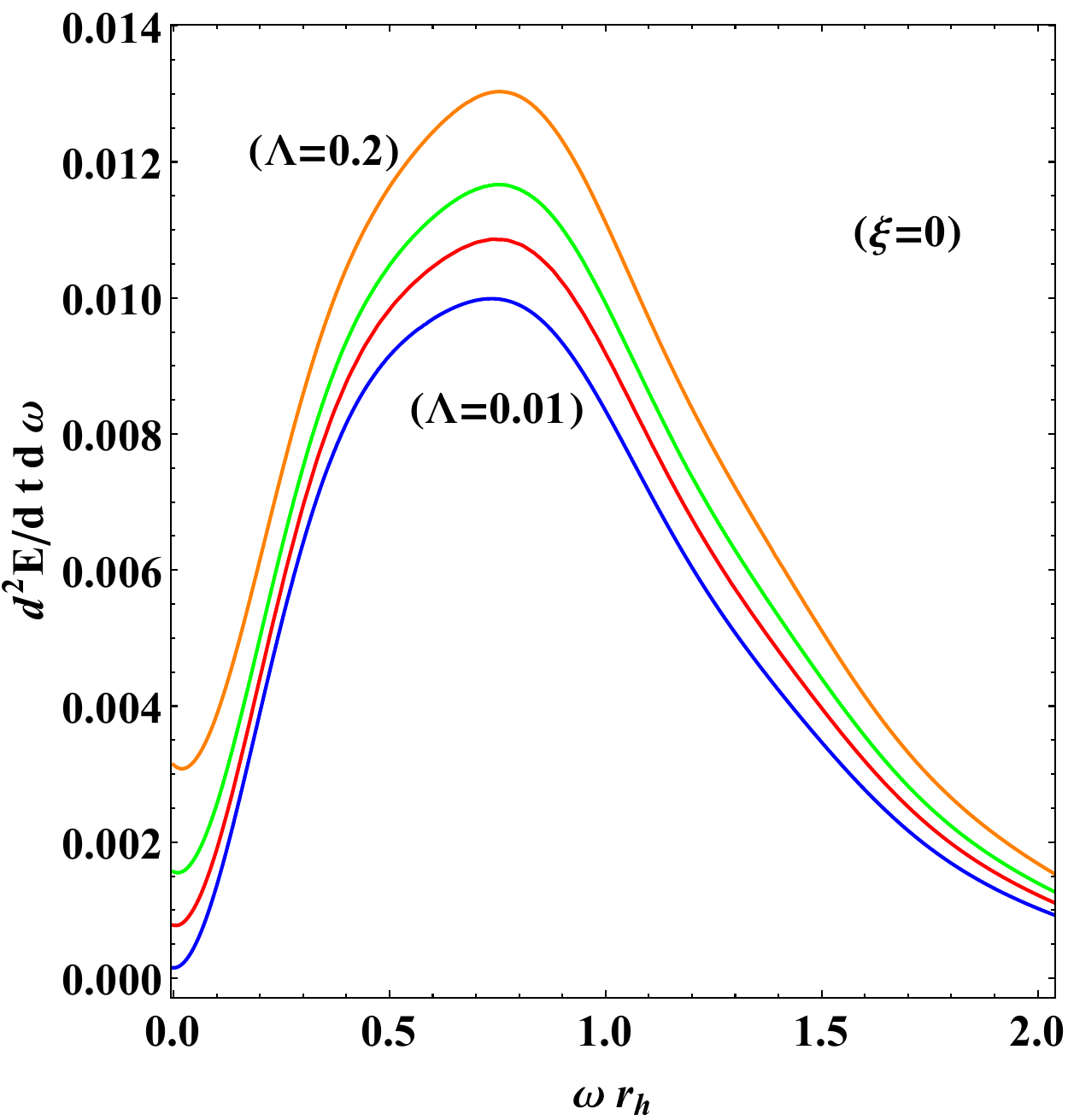}
        \caption{}
        \label{eer_br_xi00}
    \end{subfigure}
    ~ 
    \begin{subfigure}[b]{0.318\textwidth}
        \includegraphics[width=\textwidth]{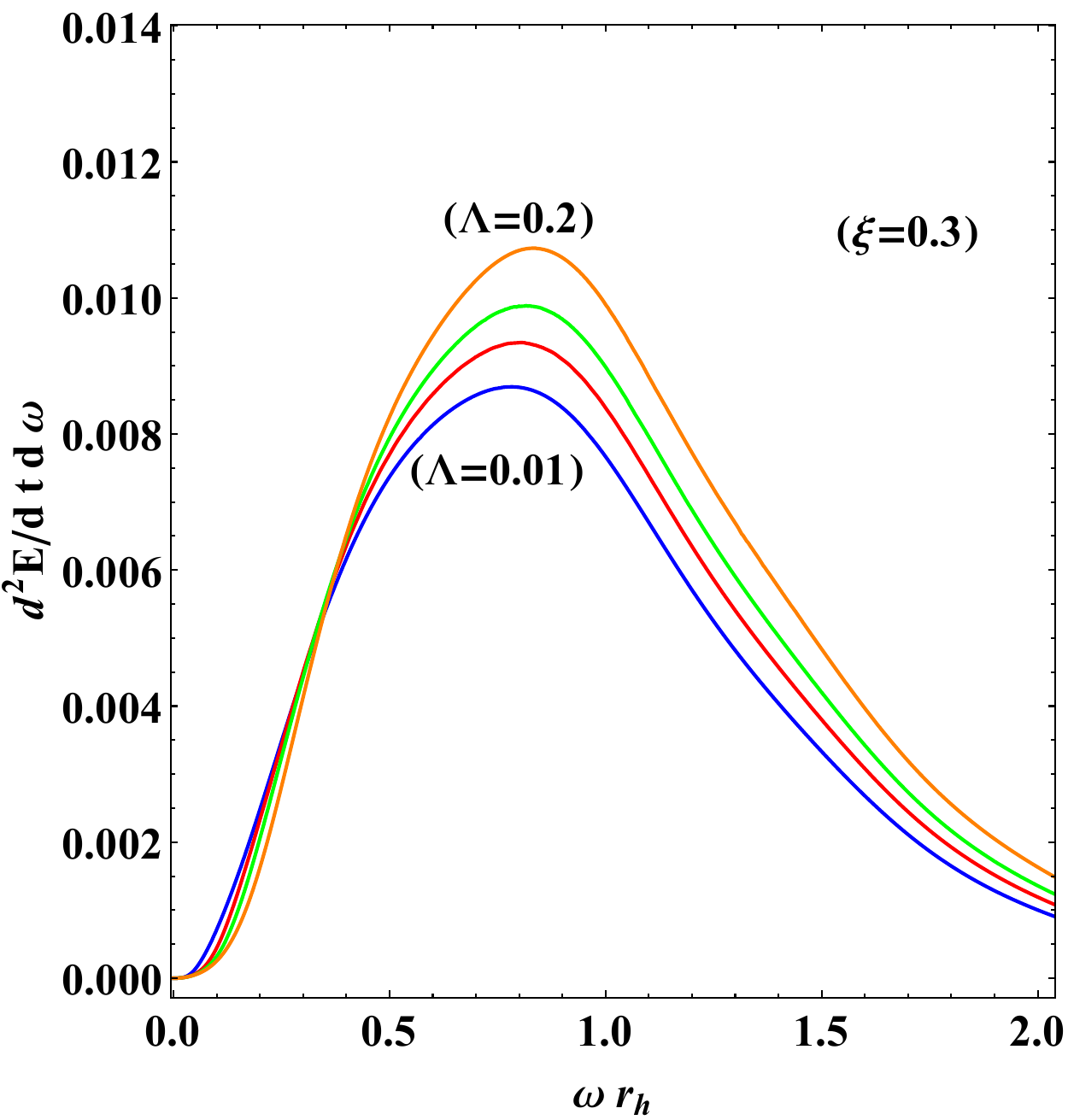}
        \caption{}
        \label{eer_br_xi03}
    \end{subfigure}
    ~ 
    \begin{subfigure}[b]{0.318\textwidth}
        \includegraphics[width=\textwidth]{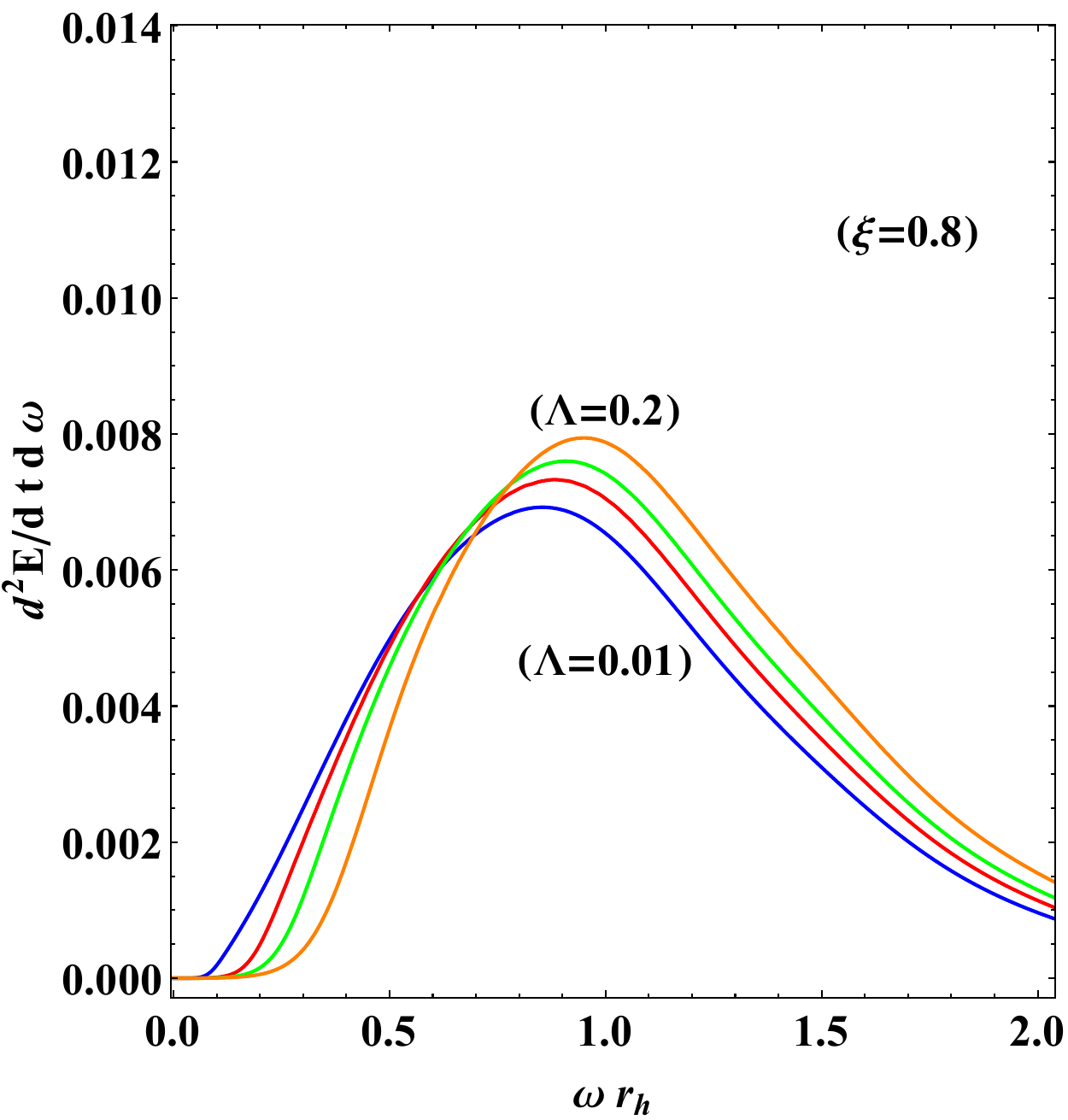}
        \caption{}
        \label{eer_br_xi08}
    \end{subfigure}
    \caption{(colour online). Energy emission rates for brane scalar fields for $n = 2$, variable
$\Lambda = 0.01, 0.05, 0.1, 0.2$, and for: (a) $\xi = 0$, (b) $\xi = 0.3$, and (c) $\xi=0.8$. }\label{eer_br_var_L}
\end{figure}

In Fig. \ref{eer_br_n_xi}(b), we also observe that for the case of minimal coupling, i.e. for $\xi=0$,
the energy emission curve starts from a non-vanishing low-energy value. This feature was first
found in \cite{KGB} and later confirmed in \cite{Harmark, Wu, Crispino}: it is attributed to the
non-vanishing, geometric, low-energy value of the greybody factor for scalar fields and leads to a 
non-zero probability for the emission of scalar particles with extremely low energy on the brane.
The same behaviour is depicted in Fig. \ref{eer_br_var_L}(a), where the emission curves for
the minimal coupling case and for various values of the cosmological constant are shown. Again, the
results derived previously in the aforementioned works are confirmed: as $\Lambda$ increases,
the energy emission rate is uniformly enhanced and the same holds for its asymptotic value
at the very low-energy regime. 

However, as soon as the non-minimal coupling constant takes non-vanishing values, the situation
changes radically also for the power spectra: the low-energy asymptotic value of the emission
rate is now zero and the dependence on the cosmological constant depends both on the value
of $\xi$ and the energy regime. For intermediate values of $\xi$, as displayed in
Fig. \ref{eer_br_var_L}(b), the enhancement in terms of $\Lambda$ appears only in the
intermediate and high-energy regime and is of a smaller magnitude. For larger values of
$\xi$, as in Fig. \ref{eer_br_var_L}(c), the emission rate is even less enhanced in the
aforementioned regimes while at the low-energy regime is actually suppressed in terms of $\Lambda$.


\subsection{Power Spectra for emission in the bulk}

The differential energy emission rate for the Hawking radiation in the form of bulk scalar
fields is again given by Eq. (\ref{diff-rate-brane}). Bulk and brane modes ``see'' the same
black-hole temperature (\ref{T_BH}), however, the multiplicity of states that correspond to
the same angular-momentum number is now different: due to the enhanced spherical symmetry
of the $(4+n)$-dimensional spacetime, $N_l$ is now given by \cite{Bander, KGB}
\beq
N_l=\frac{(2l+n+1)\,(l+n)!}{l!\,(n+1)!}\,.
\label{Nl_extra}
\eeq
Another important difference is, of course, the expression for the greybody factor $|A|^2$ that
now describes the transmission of scalar fields through the bulk potential barrier. In order to
derive the complete, exact power spectra for emission of bulk scalar fields, the numerical
results for the corresponding greybody factor found in Section 3.2 will be used. 

\begin{figure}[t!]
    \centering
    \begin{subfigure}[ht]{0.455\textwidth}
        \includegraphics[width=\textwidth]{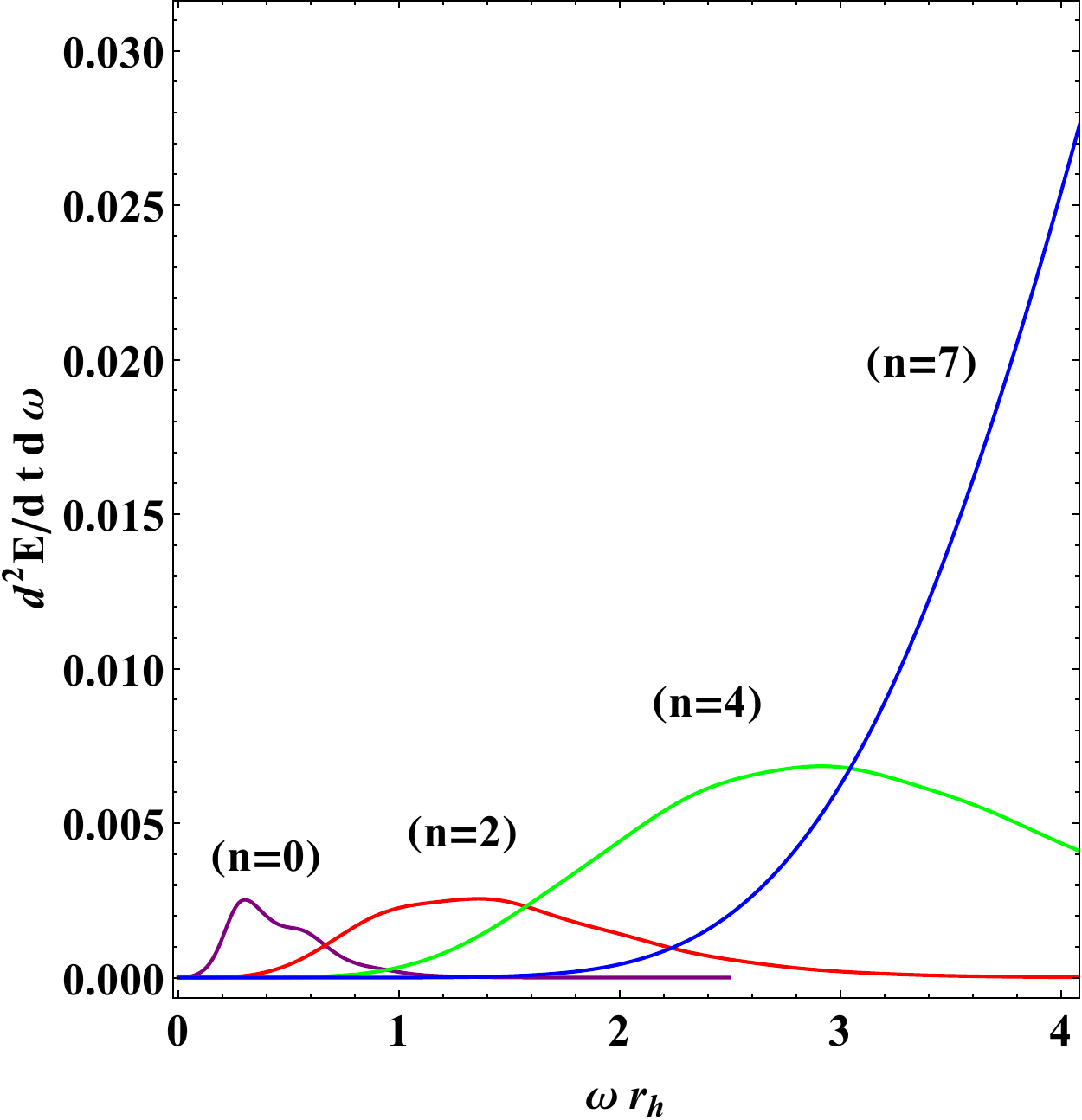}
        \caption{}
        \label{eer_bu_n}
    \end{subfigure}
    ~ 
    \begin{subfigure}[ht]{0.47\textwidth}
        \includegraphics[width=\textwidth]{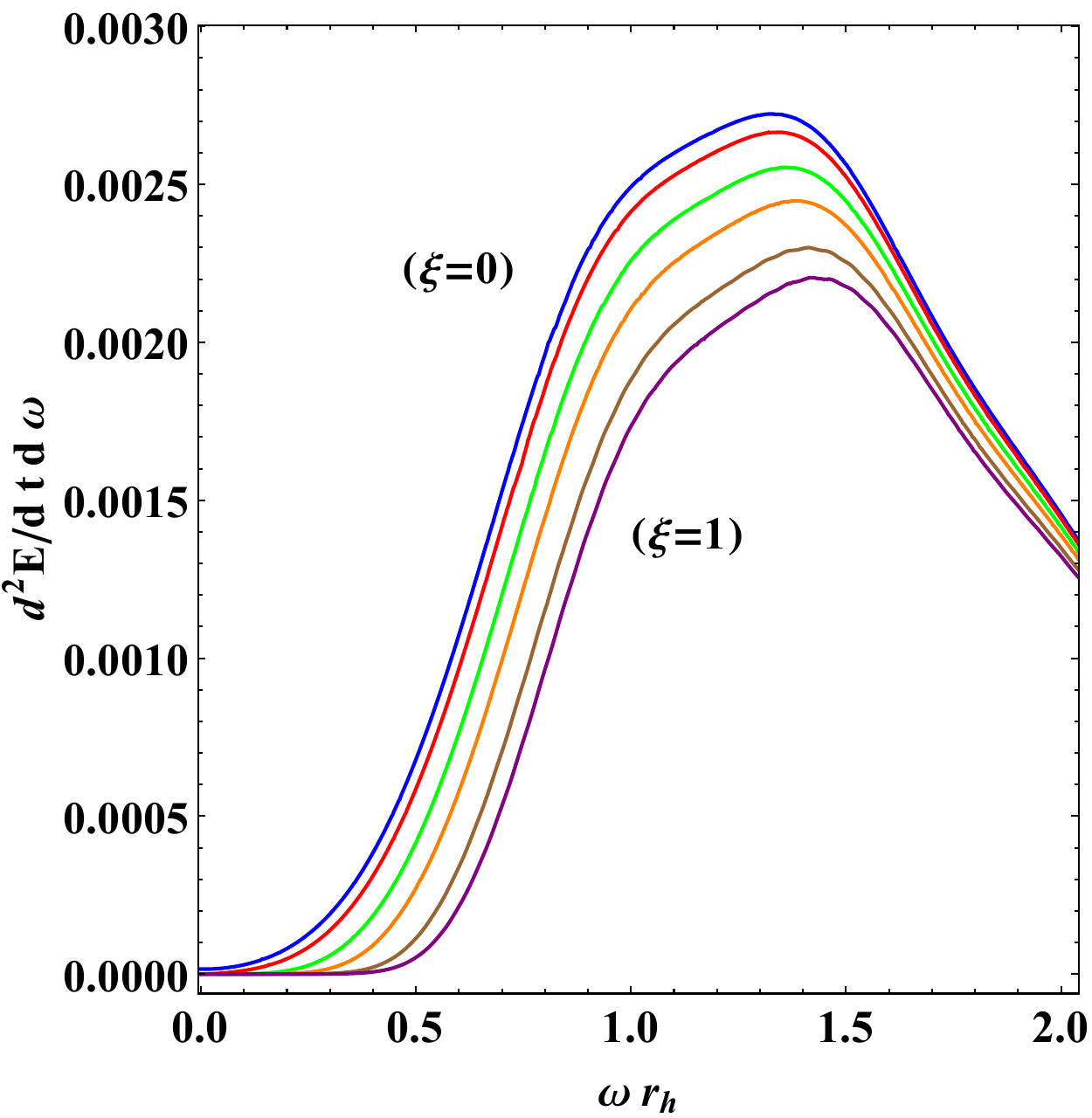}
        \caption{}
        \label{eer_bu_xi}
    \end{subfigure}
    \caption{(colour online). Energy emission rates for bulk scalar fields for $\Lambda = 0.1$,
and: (a) $\xi = 0.3$ and variable $n = 0, 2, 4, 7$; and (b) $n = 2$ and variable 
$\xi = 0, 0.1, 0.3, 0.5, 0.8, 1$.    }\label{eer_bu_n_xi}
\end{figure}

In Fig. \ref{eer_bu_n_xi}(a), we depict the dependence of the energy emission rate for scalar
fields in the bulk in terms of the number of extra dimensions. The peaks of the energy
emission curves are shifted towards larger frequencies, in a more prominent way compared
to the case of brane emission, and their heights increase leading again to a significant
enhancement with the number of extra dimensions. A comparison of Figs.
\ref{eer_br_n_xi}(a) and \ref{eer_bu_n_xi}(a) shows that the bulk emission curves lie lower
than the corresponding brane emission curves pointing to the dominance of the brane over
the bulk emission channel -- however, this holds for the particular sets of parameters chosen 
while, as we will see in the next subsection, a different set of parameters may reveal an
altogether different picture. 

The expected suppression of the power spectra in terms of the coupling constant $\xi$ is indeed
observed in Fig. \ref{eer_bu_n_xi}(b), also for the bulk scalar fields. The general profile of the emission
curves, as $\xi$ varies, is again in complete agreement with the behaviour of the energy emission
curves for bulk, massive particles \cite{KP1} as their mass changes. Note, that
in the case of minimal coupling, we should also observe a non-vanishing asymptotic emission rate
as the energy goes to zero; however, the smallness of the corresponding asymptotic value
of the greybody factor for bulk scalar fields, for the particular value of $\Lambda$ chosen
(i.e. $\Lambda=0.1$), leads to the small value
of the energy emission rate, and a zoom-in plot would be again necessary to observe this.

\begin{figure}[t!]
    \centering
    \begin{subfigure}[b]{0.318\textwidth}
        \includegraphics[width=\textwidth]{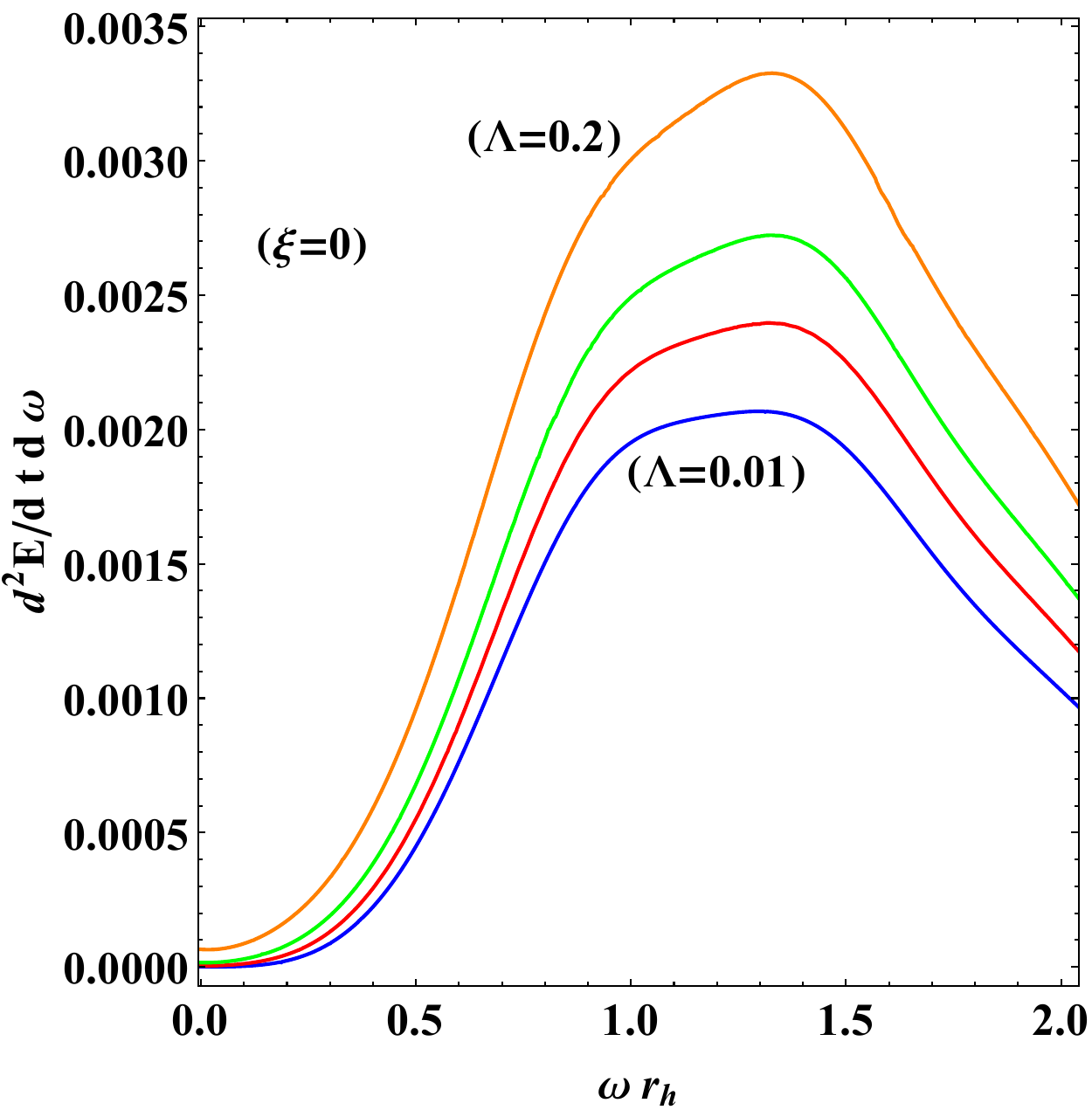}
        \caption{}
        \label{eer_bu_xi00}
    \end{subfigure}
    ~ 
    \begin{subfigure}[b]{0.318\textwidth}
        \includegraphics[width=\textwidth]{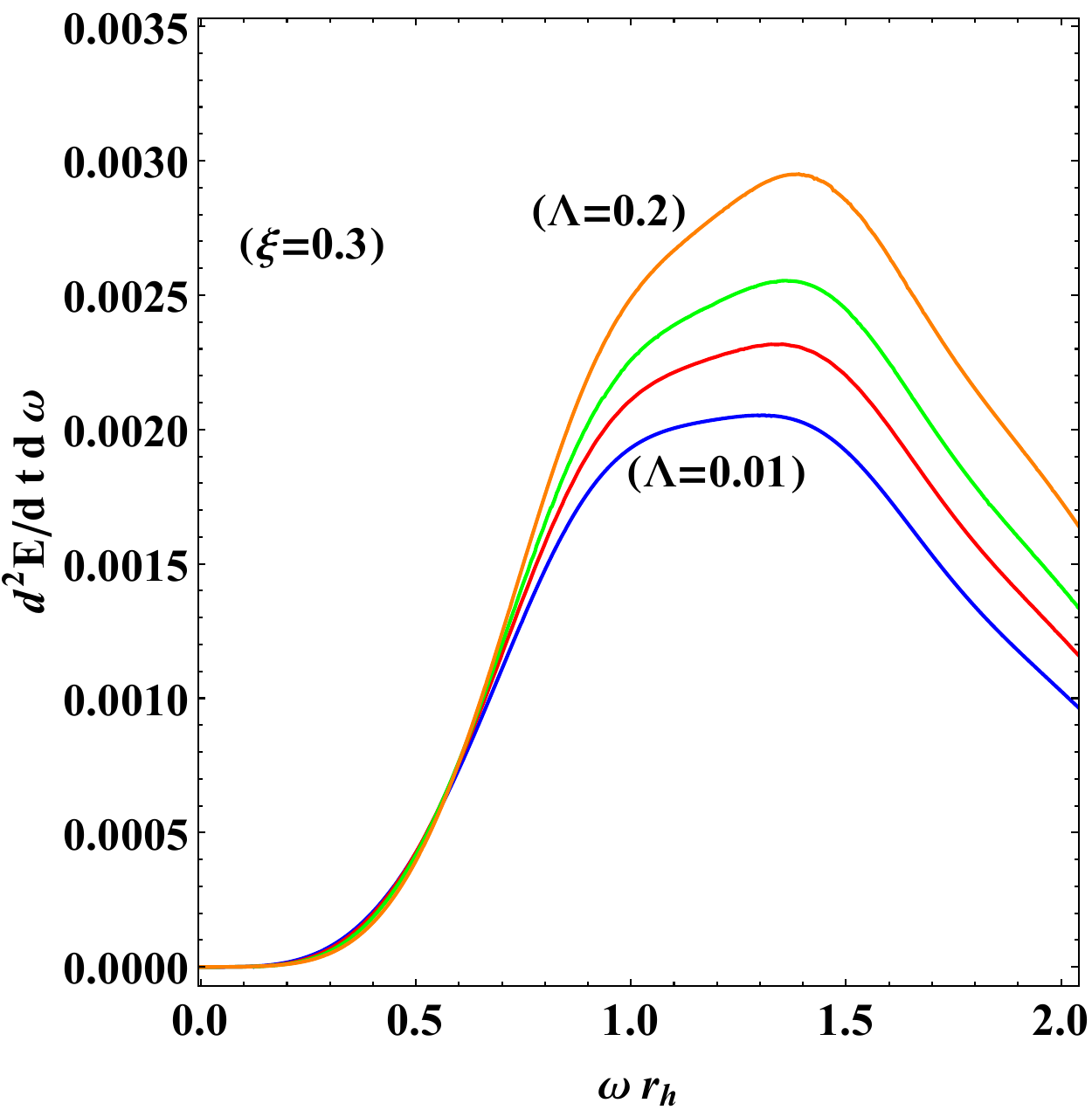}
        \caption{}
        \label{eer_bu_xi03}
    \end{subfigure}
    ~ 
    \begin{subfigure}[b]{0.318\textwidth}
        \includegraphics[width=\textwidth]{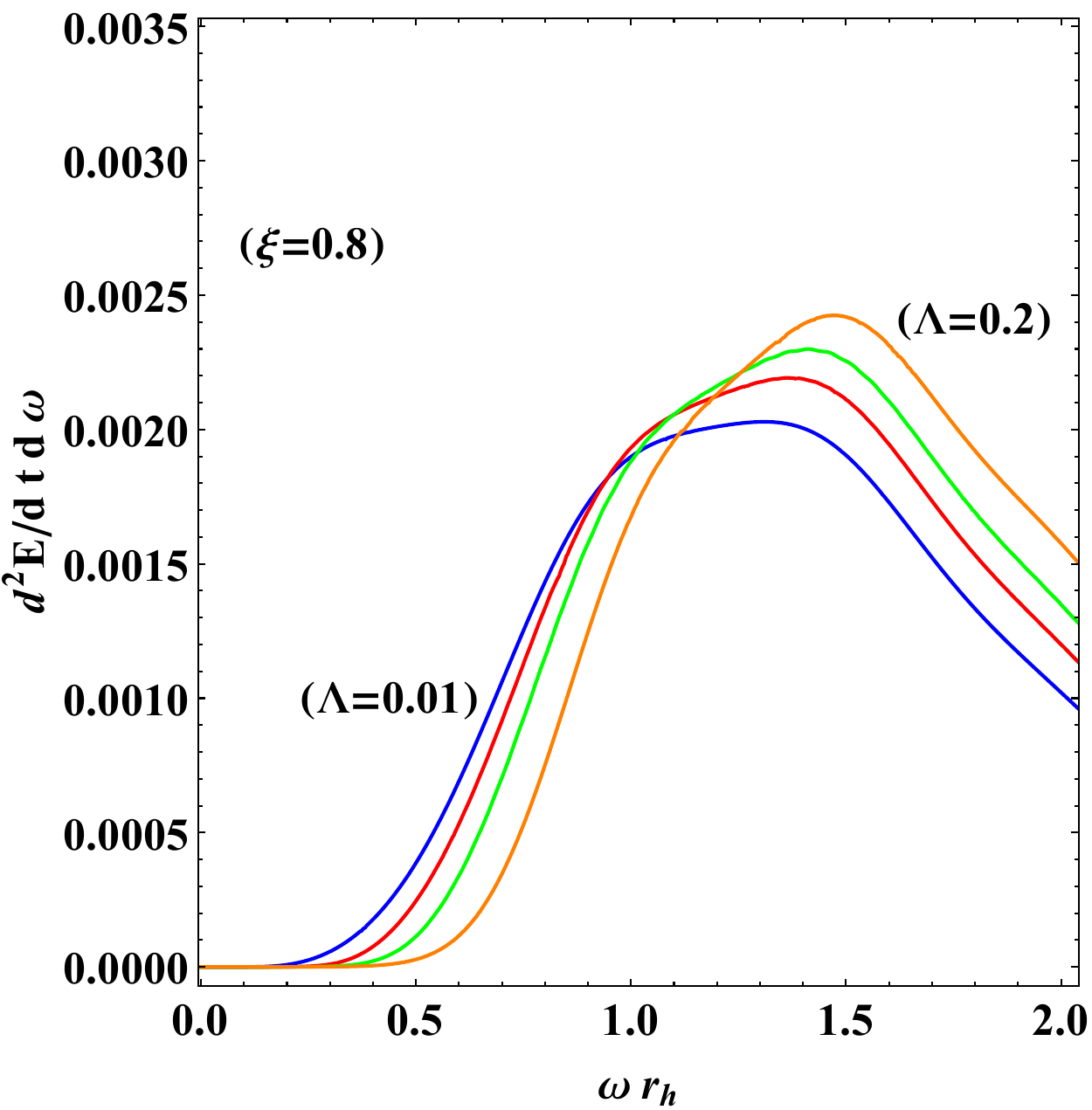}
        \caption{}
        \label{eer_bu_xi08}
    \end{subfigure}
    \caption{(colour online). Energy emission rates for bulk scalar fields for $n = 2$, variable
$\Lambda = 0.01, 0.05, 0.1, 0.2$, and for: (a) $\xi = 0$, (b) $\xi = 0.3$, and (c) $\xi=0.8$. }\label{eer_bu_var_L}
\end{figure}

This feature is more easily seen in Fig. \ref{eer_bu_var_L}(a), depicting the minimal-coupling
case with $\xi=0$, where a larger value has been chosen for the cosmological constant for the
upper curve (i.e. $\Lambda=0.2$). Also, in this case, we may see the
enhancement of the power spectra over the whole energy regime as the cosmological constant
increases. As the coupling constant takes on non-vanishing values, we observe a similar
behaviour to the one for brane emission: for moderate values of $\xi$, the enhancement
is restricted in the intermediate and high-energy regimes, while for larger values of $\xi$,
the high-energy, limited enhancement is accompanied by a low-energy suppression in 
terms of $\Lambda$ -- see Figs. \ref{eer_bu_var_L}(b,c), respectively.

\subsection{Bulk versus Brane: Relative Emission Rates and Total Emissivities}

In this final subsection, we compare the relative emission rates of the higher-dimensional 
Schwarzschild-de-Sitter black hole in the bulk and on the brane. We will demonstrate that
the parameters of this model affect significantly the amount of energy emitted on the brane
compared to the one in the bulk, and, for certain ranges of parameters, they can even tilt
the energy balance in favour of the bulk channel. 

\begin{figure}[t!]
    \centering
    \begin{subfigure}[b]{0.485\textwidth}
        \includegraphics[width=\textwidth]{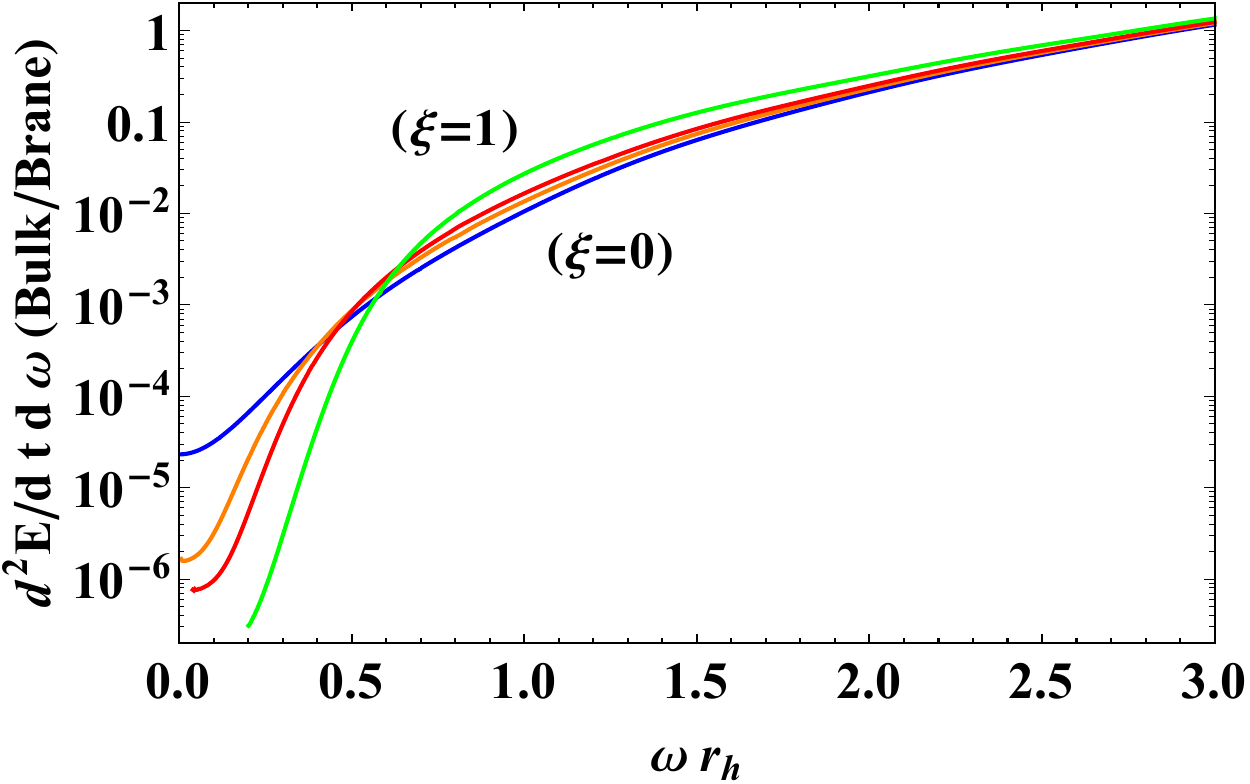}
        \caption{}
        \label{relative_var_xi}
    \end{subfigure}
    ~ 
    \begin{subfigure}[b]{0.48\textwidth}
        \includegraphics[width=\textwidth]{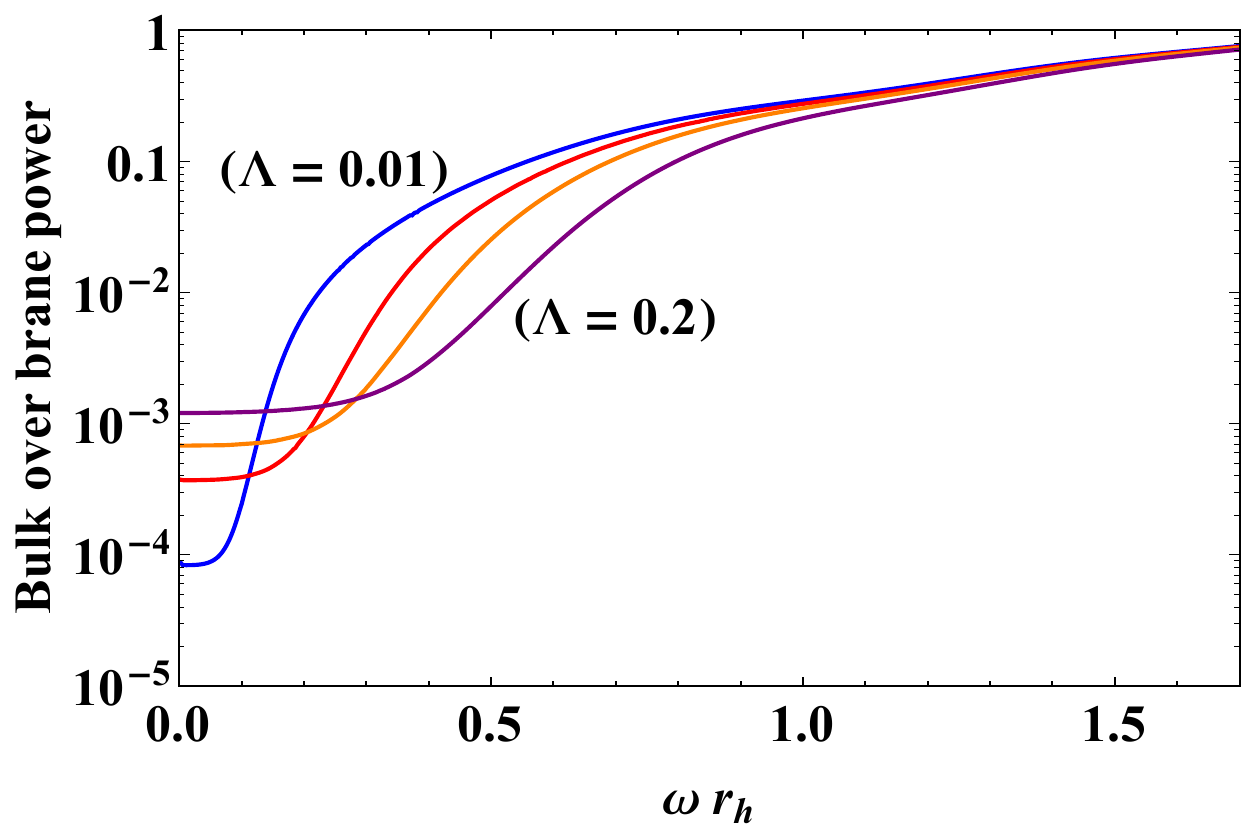}
        \caption{}
        \label{relative_var_L}
    \end{subfigure}
    \caption{(colour online). Bulk-over-brane relative emission rates for: (a) $n = 4$ ,
$\Lambda = 0.1$ and variable $\xi = 0,0.3,0.5,1$, and (b) $\xi = 0.8$, $n=2$
and variable $\Lambda = 0.01, 0.05, 0.1, 0.2$. }\label{relative_var_xi_and_L}
\end{figure}

The effect of the field coupling $\xi$ on the bulk-to-brane energy emission ratio is given in
Fig. \ref{relative_var_xi_and_L}(a), where we have fixed the number of extra dimensions and
value of cosmological constant to $n=4$ and $\Lambda = 0.1$, respectively. The coupling
constant $\xi$ assumes a dual role, depending on which part of the energy spectrum we
consider: at the low-energy regime, a large value of the coupling favours the brane emission
while beyond the intermediate regime, is the bulk emission that is now enhanced. Indeed,
an inspection of Figs. \ref{eer_br_n_xi} and \ref{eer_bu_n_xi} clearly shows that, for the same
choice of parameters, the bulk emission at the low-energy regime is very much suppressed
compared to the brane one, and this pushes the bulk-to-brane ratio to extremely small values; at
the high-energy regime, though, the same plots reveal that the two energy emission rates
differ by, at most, an order to magnitude or even less depending on the values of $n$, 
$\Lambda$ and $\omega r_h$. According to Fig. \ref{relative_var_xi_and_L}(a), the
bulk-to-brane ratio remains smaller than unity but it is clearly pushed to values close to,
or even above, unity at the high-energy regime. This is again in agreement with the role
that the mass of a scalar field plays in the bulk-to-brane ratio: as was shown in \cite{KP1},
the mass of the scalar field gives a significant boost to the bulk channel.

According to the behaviour depicted in Fig. \ref{relative_var_xi_and_L}(b), the cosmological
constant plays a similar dual role: at the very low-energy regime, an increase in its value
seems to enhance the bulk channel, at the intermediate regime it is the brane channel
that is enhanced instead, while at the high-energy regime the bulk-to-brane
ratio becomes almost insensitive to the value of the cosmological constant. This change
of role is due to the presence of the coupling constant which here has been
given the value $\xi=0.8$; in the minimal coupling case, studied in \cite{KGB}, the
increase in the value of $\Lambda$ led to the enhancement of the bulk channel
throughout the low and intermediate-energy regime. Again, for the range of parameters
shown in this plot, the energy ratio seems to approach unity as the energy parameter
increases further, and, as a result, highly-energetic particles have equal emission
rates on the brane and in the bulk. 

\begin{figure}[t!]
  \begin{center}
\mbox{\includegraphics[width = 0.6 \textwidth] {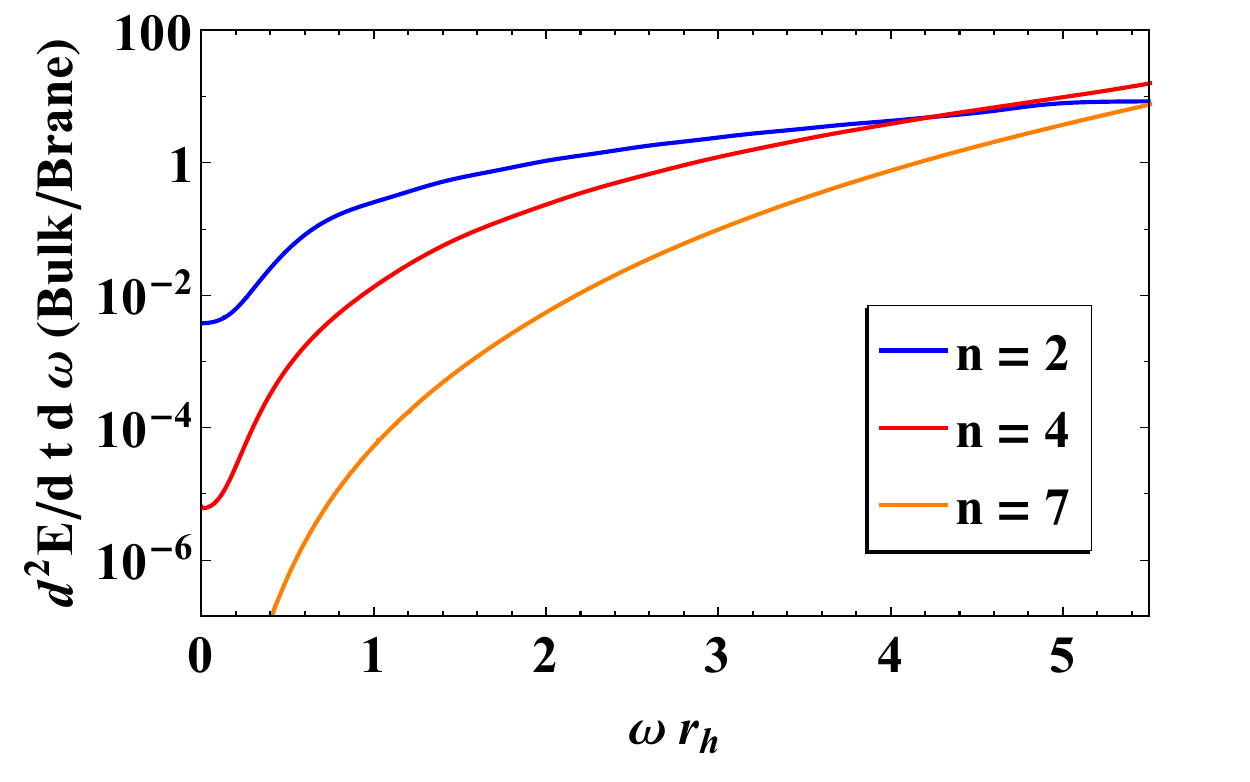}}
  \caption{(color online). Bulk-over-brane relative emission rates for $\xi = 0.3$, $\Lambda = 0.2$ and variable $n = 2, 4, 7$.}
   \label{relative_var_n}
  \end{center}
\end{figure}

Last, but not least, we should address also the effect that the number of extra dimensions
has on the value of the bulk-to-brane ratio. In \cite{KGB}, where the minimal coupling
case was studied, it was demonstrated that, up to the intermediate energy regime, the
bulk-to-brane ratio remained below unity, and it actually decreased as $n$ increased. 
This is the behaviour we obtain also in Fig. \ref{relative_var_n}, where the coupling constant
has now adopted a moderate, non-vanishing value ($\xi=0.3$). At the low and
intermediate-energy regime, there is a clear brane emission domination; however, at larger
frequencies the curves for different values of $n$ cross each other revealing an enhancement
in the emission of bulk scalars over brane scalars with the number of extra dimensions.
Finally, at high energies, the bulk emission is clearly favored over the brane one for all
values of $n$, and the bulk channel dominates.

\begin{figure}[b!]
\begin{center}
\centering
    \begin{subfigure}[b]{0.47\textwidth}
        \includegraphics[width=\textwidth]{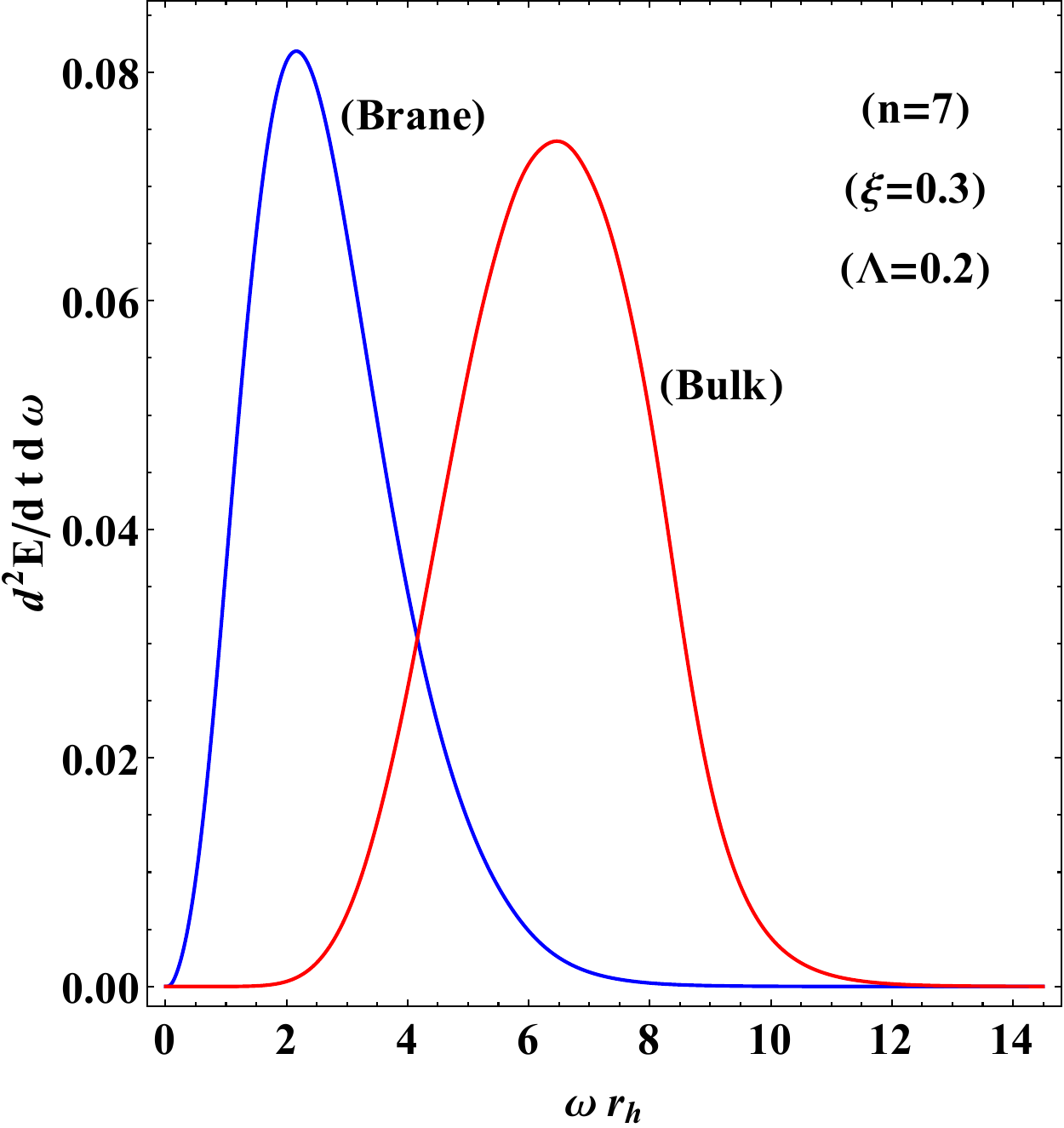}
        \caption{}
    \end{subfigure}
    ~ 
    \begin{subfigure}[b]{0.47\textwidth}
        \includegraphics[width=\textwidth]{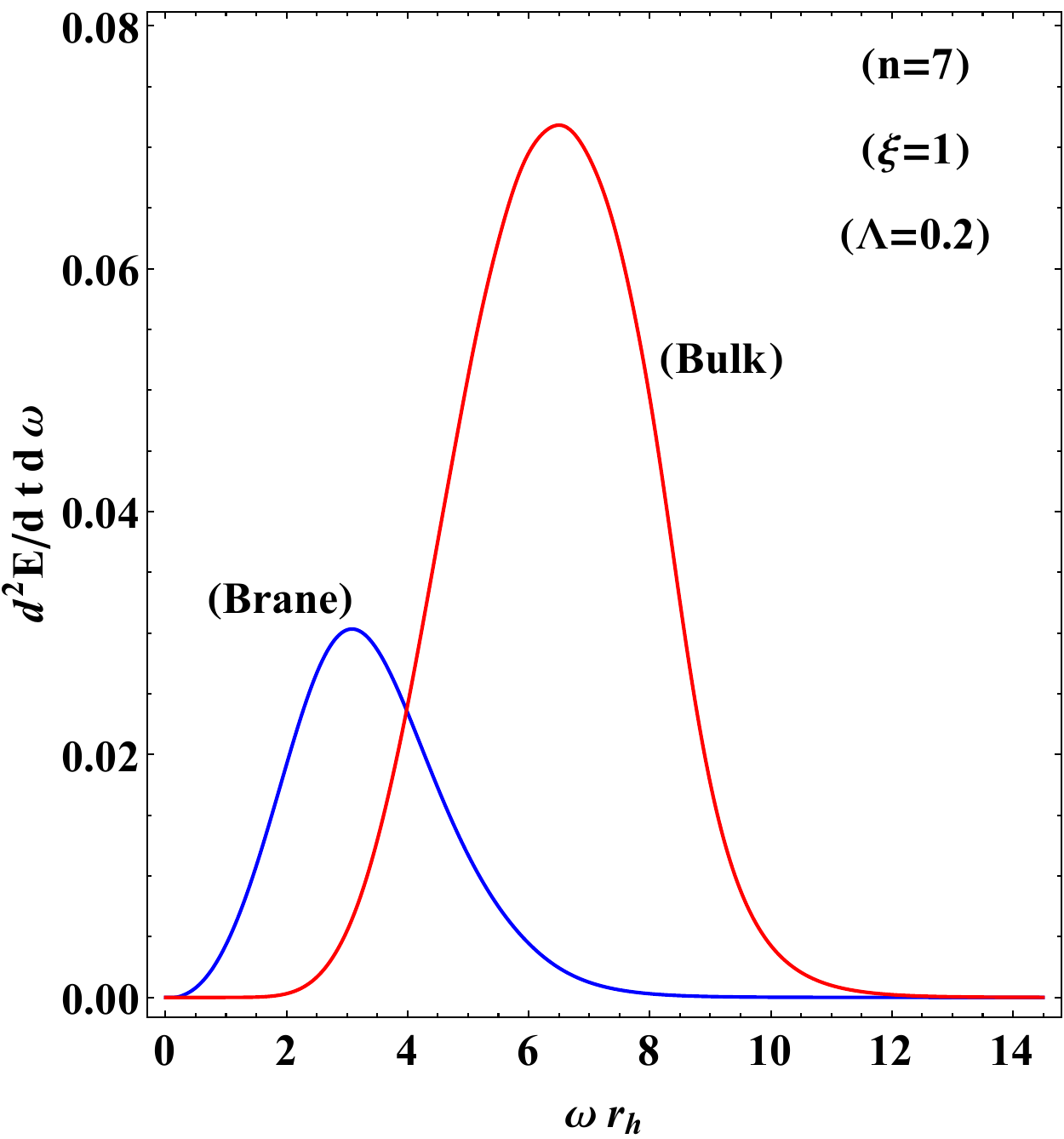}
        \caption{}
    \end{subfigure}
\caption{(color online). Power spectra for emission on the brane and in the bulk for $n = 7$,
$\Lambda=0.2$, and: (a) $\xi = 0.3$, and (b) $\xi=1$.}
   \label{bulk_over_brane_eer_domination}
  \end{center}
\end{figure}

As an indicative example of the combined effect that the number of extra dimensions, cosmological
constant and non-minimal coupling may have on the emission rates, in 
Figs. \ref{bulk_over_brane_eer_domination}(a,b), we depict the bulk and brane emission curves
for $n=7$, $\Lambda=0.2$ and for two different values of the coupling constant, $\xi=0.3$
and $\xi=1$, respectively. For the former value of $\xi$, Fig. \ref{bulk_over_brane_eer_domination}(a)
justifies several of the features already discussed in Fig. \ref{relative_var_n}: for this set of parameters,
the brane emission is indeed dominant at the low-energy regime but the bulk dominates in the
emission of high-energy modes. As $\xi$ adopts the latter value, Fig. \ref{bulk_over_brane_eer_domination}(b)
reveals a picture with the same qualitative profile but with a fundamental difference from the
quantitative point of view: the dominance of the bulk emission is now so strong that the emission
on the brane seems to comprise only a small part of the total black-hole emission.

The aforementioned behaviour makes imperative the calculation and comparison of the brane
and bulk total emissivities, i.e. the amount of energy emitted by the black hole on the brane
and in the bulk in the unit of time over the whole frequency range. To this end, we integrate
the differential energy emission rate (\ref{diff-rate-brane}) over
the frequency $\omega$; this is equivalent to computing the area under the energy emission
curves. We perform this for a range of values of our parameters, and at the end we compute
the bulk-to-brane ratio of emissivities; this ratio will be a good index of how the energy balance
between brane and bulk changes in terms of the parameters. The obtained values are displayed in
Tables \ref{table:totaln2} through \ref{table:totaln7}.

\begin{table}[t!]
\caption{Bulk over brane total emissivity for $n=2$} 
\centering 
\begin{tabular}{|c || c| c| c| c| c| c|} 
\hline\ 
$  \xi \rightarrow$ & 0.0 & 0.1 & 0.3 & 0.5 & 0.8 & 1.0 \\ [0.5ex] 
\hline\hline 
$\Lambda = 0.01$ & 0.257506 & 0.269481 & 0.294391 & 0.320639 & 0.362918 & 0.393068   \\ 
\hline $0.05$ & 0.27356 & 0.285502 & 0.309271 & 0.333195 & 0.369824 & 0.394932  \\
\hline$0.1$ & 0.288635 & 0.300295  & 0.322187 & 0.343  & 0.373032 & 0.392523  \\
\hline$0.2$ & 0.314566 & 0.325492 & 0.343106 & 0.357599 & 0.375749 & 0.38618  \\ 
\hline 
\end{tabular}
\label{table:totaln2} 
\end{table}

\begin{table}[t!]
\caption{Bulk over brane total emissivity for $n=4$} 
\centering 
\begin{tabular}{|c || c| c| c| c| c| c|} 
\hline\ 
$  \xi \rightarrow$ & 0.0 & 0.1 & 0.3 & 0.5 & 0.8 & 1.0 \\ [0.5ex] 
\hline\hline 
$\Lambda = 0.01$ & 0.247028 & 0.275432 & 0.339321 & 0.413966 & 0.549627 & 0.658426   \\ 
\hline $0.05$ & 0.255885 & 0.284767 & 0.349056 & 0.423319 & 0.556669 & 0.662449   \\
\hline$0.1$ & 0.264557 & 0.293826 & 0.358134 & 0.43146 & 0.561241 & 0.662884  \\
\hline$0.2$ & 0.279594 & 0.30942 & 0.373259 & 0.444156 & 0.566328 & 0.659702  \\ 
\hline 
\end{tabular}
\label{table:totaln4} 
\end{table}

\begin{table}[t!]
\caption{Bulk over brane total emissivity for $n=7$} 
\centering 
\begin{tabular}{|c || c| c| c| c| c| c|} 
\hline\ 
$  \xi \rightarrow$ & 0.0 & 0.1 & 0.3 & 0.5 & 0.8 & 1.0 \\ [0.5ex] 
\hline\hline 
$\Lambda = 0.01$ & 0.779006 & 0.906992 & 1.21479 & 1.60427 & 2.38062 & 3.05751   \\ 
\hline $0.05$ & 0.790883 & 0.92003 & 1.22955 & 1.61993 & 2.39509 & 3.0686  \\
\hline$0.1$ & 0.803103 & 0.933293 & 1.24413 & 1.63466 & 2.40657 & 3.07444  \\
\hline$0.2$ & 0.824629 & 0.956511 & 1.26906 & 1.65866 & 2.42208 & 3.07722  \\ 
\hline 
\end{tabular}
\label{table:totaln7} 
\end{table}


Starting from the minimal-coupling case with $\xi=0$, we observe that, independently of the
value of $\Lambda$, the bulk-to-brane emissivity ratio takes a dip for intermediate values
of $n$, and then rises again as $n$ takes on larger values; this feature was observed also
for the higher-dimensional Schwarzschild black hole \cite{HK1} and, according to our results,
persists also in the case of a Schwarzschild-de-Sitter background. As $\Lambda$ ranges 
from 0.01 to 0.2, for fixed $n$, the bulk-to-brane ratio is enhanced but by a moderate
amount - we estimate that, for small values of $n$, by merely increasing the value of the
cosmological constant we would not obtain a ratio larger than unity, and thus a bulk
domination. The same seems to hold for small values of $n$ and $\Lambda$ in terms of
the coupling $\xi$: a moderate enhancement appears as $\xi$ changes from 0 to 1 
(in fact, for $\xi=1$, a slight decrease appears in the ratio as $\Lambda$ increases,
a feature that is probably attributed to the slightly more significant suppression of
the bulk emission curves compared to the brane ones at the low-energy for this set
of parameters).  

As $n$ increases, though, from $n=2$ to $n=4$ and finally to $n=7$, the enhancement
in terms of $\xi$ becomes significant reaching a factor of order 3 or 4. This enhancement
is again justified by the different suppression the effective mass has on the energy emission
rates: for a brane scalar field the effective mass comes out to be larger than the one for
a bulk scalar field by a factor of ${\cal O}(10)$, which then causes the significant suppression
of the emission of the corresponding brane fields. The enhancement factor in terms of $\xi$,
for the case $n=7$, is now enough to raise the bulk-to-brane ratio to values larger than
unity: for $\xi=1$, the energy emitted in the bulk is 3 times larger than the one emitted
on the brane. We expect that this ratio will take even larger values as $\xi$ is increased
further.


\section{Conclusions}

In this work, we have studied the problem of propagation of scalar fields in the gravitational
background of a higher-dimensional Schwarzschild-de-Sitter black hole as well as on
the projected-on-the-brane 4-dimensional background. The scalar fields were also
assumed to have a non-minimal coupling to the corresponding, brane or bulk, scalar
curvature that effectively acted as a mass term. Previous studies had
addressed the topic of only minimally-coupled scalar fields in a higher-dimensional
background \cite{KGB, Harmark, Wu} or non-minimally coupled fields in a 4-dimensional
Schwarzschild-de-Sitter background \cite{Crispino}. A recent work of ours \cite{KPP1}
had attacked the full problem
but in an analytic way: that approach allowed us to derive analytic expressions for the
bulk and brane scalar greybody factors and to study their properties; however, the
validity of the approximate technique used relied on the assumption that both the
cosmological constant and the non-minimal field coupling were small. 

Here, we have returned to the same problem and performed a comprehensive study by
deriving exact numerical results for propagation of non-minimally-coupled scalar fields
both in the bulk and on the brane. We have dealt with each case separately, solved
the corresponding equations of motion numerically in order to derive the radial part
of the field and, in terms of the latter, determined the greybody factors, or transmission
probabilities. The analytical solutions we had previously derived served as asymptotic
boundary conditions as well as checking points for our numerical analysis. Both in
the bulk and on the brane, we demonstrated that, as expected, there was a very good 
agreement between the analytic and numerical results at the low-energy regime, for
small values of the cosmological constant and the non-minimal field coupling, while deviations
started to appear as $\Lambda$, $\xi$ or the energy of the emitted particle increased
beyond the allowed regimes.

Our exact, numerical analysis allowed us to study in detail the behaviour of the greybody
factors in terms of all the parameters of the model: the angular-momentum number $l$ and
non-minimal coupling $\xi$ of the field, the cosmological constant $\Lambda$  and number
$n$ of extra spacelike dimensions
of spacetime. We have confirmed the suppression of the greybody factor, both for brane
and bulk scalar fields, as either $l$ or $\xi$ increased. The same holds for the number
of extra dimensions, while the dependence on the cosmological constant proved to be
more subtle, exactly as our previous analytic study had hinted to: depending on the
value of the non-minimal coupling constant, $\Lambda$ can either help the emitted particle
to overcome the gravitational barrier or suppresses its transmission probability.  

Moving beyond the boundaries of our previous analytic work, and having at our disposal
the exact greybody factors for arbitrary values of the parameters of the model,
we then proceeded to calculate the differential energy emission rates for the Hawking radiation
in the form of scalar fields by a higher-dimensional Schwarzschild-de-Sitter black hole both
in the bulk and on the brane. The behaviour of the power spectra in terms of the parameters
of the model was the result of the contributions of both the greybody factors and the
black-hole temperature. Similarly for brane and bulk emission, the differential energy
rates received non-negligible contributions only from the first six partial modes due
to the spherical symmetry of the background. An increase in the number of extra dimensions,
due to the significant enhancement of the black-hole temperature, resulted to the 
enhancement itself of the emission rate, both on the brane and in the bulk. The
non-minimal coupling term of the scalar field with the -- brane or bulk -- scalar curvature
acted as an effective mass term, as anticipated, with any increase in the value of the
coupling $\xi$ causing the suppression of the emission rates throughout the energy
regime. The more subtle role of the cosmological constant was reflected also in the
profile of the power spectra: for small values of $\xi$, the emission rate is enhanced
at all frequencies, while as $\xi$ increases the enhancement persists only at the
high-energy regime and a simultaneous suppression appears at the lower part of the
spectrum.

A conclusion that was also drawn from the above analysis was that the greybody factors
for brane and bulk scalar fields exhibited differences in their profiles in terms of the
parameters of the model; these differences appeared also in the corresponding energy
emission rates. By computing the relative energy rates, we showed that, for small
values of the cosmological constant and field coupling, the bulk channel was 
sub-dominant at the low and intermediate energy regime; as noted before,
the bulk channel was significantly enhanced at the high-energy regime without
surpassing though the brane channel. However,
as the non-minimal field coupling, or equivalently the effective mass term, of the
scalar fields increased, the bulk channel received an extra boost. When the above
was combined also with a large number of extra spacelike dimensions, the bulk
channel became the dominant one. The calculation of the exact ratio of bulk and
brane emissivities showed that for $n=7$ and $\xi=1$ - a rather moderate value of the
field coupling - the bulk channel emitted more energy per unit time by a factor of
almost 4 compared to the brane channel. This has been one of the very few times
(see also \cite{GBK}) where the brane domination not only breaks down but, for a
further increase of the field coupling, the brane emission could be only a very small
part of the total output of energy from the Schwarzschild-de-Sitter black hole - and
this has been caused by the addition of a legitimate interaction term of the fields
under study with the scalar curvature of spacetime.


{\bf Acknowledgements.} T.P. would like to thank Dimitrios Karamitros for useful discussions.
Part of this work was supported by the COST Action MP1210 ``The String Theory Universe''.



\end{document}